# MUTUALLY ASSURED DEREGULATION

*Gilad Abiri*[*]

*We have convinced ourselves that the way to make AI safe is to make it unsafe. Since 2022, many policymakers worldwide have embraced the "Regulation Sacrifice"—the belief that dismantling safety oversight will somehow deliver security through AI dominance. This reasoning follows a perilous pattern: fearing that one major power will dominate the AI landscape at the expense of the others, we rush to eliminate any safeguard that might slow our progress. This Article reveals the fatal flaw in such thinking. Though AI development certainly poses national security challenges, the solution demands stronger regulatory frameworks, not weaker ones. A race without guardrails does not build competitive strength—it breeds shared danger.*

*The Regulation Sacrifice makes three promises. Each one is false. First, it promises durable technological leads. But as a form of dual-use software, AI capabilities spread like wildfire. Performance gaps between U.S. and Chinese systems collapsed from nine to two percent in thirteen months. When advantages evaporate in months, sacrificing permanent safety for temporary speed makes no sense.*

*Second, it promises that deregulation accelerates innovation. The opposite is quite often true. Companies report that well-designed governance frameworks streamline their development. Investment flows toward regulated markets, not away from them. Clear rules reduce uncertainty. Uncertain liability creates paralysis. We have seen this movie before—environmental standards did not kill the auto industry; they created Tesla and BYD.*

*Third, the promise of enhanced national security through deregulation is perhaps the most dangerous fallacy, as it*

---

[*] Associate Professor of Law, Peking University School of Transnational Law; Affiliate Faculty, Information Society Project, Yale Law School; and Senior Research Affiliate, Singapore Management University Centre for Digital Law.

*undermines security across timeframes. In the near term, it hands our adversaries perfect tools for information warfare. In the medium term, it puts bioweapon capabilities in everyone's hands. In the long term, it guarantees we'll deploy AGI systems we cannot control, racing to be the first to push a button we cannot un-push.*

*The Regulation Sacrifice persists because it serves powerful interests, not because it serves security. Tech companies prefer freedom to accountability. Politicians prefer simple stories to complex truths. Together, they are trying to convince us that recklessness is patriotism. But here is the punchline: these ideas create a system of mutually assured deregulation, where each nation's sprint for advantage guarantees collective vulnerability. The only way to win this game is not to play.*

INTRODUCTION

We are locked in a contest over how artificial intelligence should be governed, but we have been misled about what true victory entails.[1] Since 2022, AI policy discussions worldwide have been overtaken by a competitive framework that recasts AI development as a zero-sum contest in which regulatory restraint equals patriotic duty.[2] President Trump dismissed AI safeguards

[1] Last year, Congress considered an unprecedented ten-year moratorium on state-level AI regulation, with proponents warning that without it, the United States would lose the AI arms race to China. Although the measure was not enacted, it represents the apotheosis of a dangerous and influential paradigm that has swept from Washington across the global technology landscape. *See* H.R. 1, 119th Cong. § 43201(c)(1) (as passed by House, May 22, 2025); Matt Brown & Matt O'Brien, *House Republicans Include a 10-Year Ban on US States Regulating AI in 'Big, Beautiful' Bill*, ASSOCIATED PRESS (May 16, 2025), https://apnews.com/article/ai-regulation-state-moratorium-congress-39d1c8a0758ffe0242283bb82f66d51a [https://perma.cc/X5ZS-7F6E]; *Winning the AI Race: Strengthening U.S. Capabilities in Computing and Innovation: Hearing Before the S. Comm. on Com., Sci., & Transp.*, 119th Cong. (2025), https://www.commerce.senate.gov/2025/5/winning-the-ai-race-strengthening-u-s-capabilities-in-computing-and-innovation_2 [https://perma.cc/WDM9-RQQ4] (announcement of Sen. Ted Cruz, Chairman, S. Comm. on Com., Sci., & Transp.) [hereinafter *Winning the AI Race*].

[2] ASIA PAC. TASK FORCE, *The Age of AI in U.S.-China Great Power Competition: Strategic Implications, Risks, and Global Governance*, BEYOND THE HORIZON (Feb 3, 2025), https://behorizon.org/the-age-of-ai-in-u-s-china-great-power-competition-strategic-implications-risks-and-global-governance/ [https://perma.cc/2PQZ-7DKD]; Brian J. Chen, *The "Washington Effect" Could Decide the AI Race*, PROJECT SYNDICATE (June 19, 2025), https://www.project-syndicate.org/commentary/us-trump-administration-tries-bullying-its-way-to-ai-dominance-by-brian-j-chen-2025-06 [https://perma.cc/7MTD-XQ3V] (documenting U.S. pressure on allies to adopt deregulatory approaches); Angela Huyue Zhang, *Agility over Stability: China's Great Reversal in Regulating the Platform Economy*, 63 HARV. INT'L L.J. 457, 508–09 (2022) (arguing that China used platform-economy enforcement to push big tech toward core technologies, including AI and driverless cars); *Communication from the Commission to the European Parliament, the Council, the European Economic and Social Committee and the Committee of the Regions: AI Continent Action Plan*, at 1, COM (2025) 165 final (Apr. 9, 2025) (declaring its commitment to "becom[ing] a global leader in Artificial Intelligence" while emphasizing the need to enhance its competitiveness); *Statement on Inclusive and Sustainable Artificial Intelligence for People and the Planet at the AI Action Summit*, PERMANENT MISSION FR. UNITED NATIONS N.Y. (Feb. 11, 2025), https://onu.delegfrance.org/statement-on-inclusive-and-sustainable-artificial-intelligence-for-people-and [https://perma.cc/2YFE-3ZJQ]; Anu Bradford, *The False Choice Between Digital Regulation and Innovation*, 119 NW. U.L. REV. 377, 381 (2024) (describing how fears of losing technological leadership in a U.S.-China "tech war" fuel arguments that stricter digital regulation would harm national power and competitiveness); Anu Bradford, Eileen Li & Matthew C. Waxman, *How Domestic Institutions Shape the Global Tech War*, 16 HARV. NAT'L SEC. J. 76, 101 (2025) (explaining how national security-framed competition over advanced technologies, including AI, is used to argue that regulation undermines competitiveness and thus national security); ANU BRADFORD, DIGITAL EMPIRES: THE GLOBAL BATTLE TO REGULATE TECHNOLOGY (2023) 6-11

as threats to "American technological leadership," while threatening to withdraw from international AI agreements and pressuring allies to adopt similar deregulatory stances.[3] At the 2025 Paris AI summit, Vice President Vance warned that "excessive regulation of the AI sector could kill a transformative industry just as it's taking off," even as he cautioned that "some authoritarian regimes have stolen and used AI to strengthen their military, intelligence, and surveillance capabilities; capture foreign data; and create propaganda to undermine other nations' national security."[4] Major AI companies echo this rhetoric globally:[5] OpenAI warns of "a Chinese Communist Party determined to overtake us by 2030," while Google describes the current situation as a "global AI competition" where "policy decisions will determine the outcome."[6] The message is clear: in the sprint for AI dominance, regulation is a disastrous hurdle.

---

(framing the United States, the European Union, and China as rival "digital empires" competing to set global rules for the digital economy, including AI); Filippo Lancieri, Laura Edelson & Stefan Bechtold, *AI Regulation: Competition, Arbitrage & Regulatory Capture* 26 (Geo. U.L. Ctr. Research Paper No. 2025/05, 2025) http://dx.doi.org/10.2139/ssrn.5049259 (predicting a largely fragmented global AI regime with occasional harmonization and cautioning against race-to-the-bottom dynamics).

[3] Exec. Order No. 14,179, 90 Fed. Reg. 8741; *Fact Sheet: President Donald J. Trump Takes Action to Enhance America's AI Leadership*, THE WHITE HOUSE (Jan. 23, 2025), https://www.whitehouse.gov/fact-sheets/2025/01/fact-sheet-president-donald-j-trump-takes-action-to-enhance-americas-ai-leadership/ [https://perma.cc/EQ7B-SZHF].

[4] J.D. Vance, *Remarks by the Vice President at the Artificial Intelligence Action Summit in Paris, France*, THE AMERICAN PRESIDENCY PROJECT (Feb. 11, 2025), https://www.presidency.ucsb.edu/documents/remarks-the-vice-president-the-artificial-intelligence-action-summit-paris-france [https://perma.cc/GHH9-NWKA].

[5] *See* Amanda Levendowski, *How Copyright Law Can Fix Artificial Intelligence's Implicit Bias Problem*, 93 WASH. L. REV. 579, 612–16 (2018) (analyzing how tech companies invoke innovation arguments to resist bias mitigation requirements); Ari Ezra Waldman, *Power, Process, and Automated Decision-Making*, 88 FORDHAM L. REV. 613, 621, 628–31, 630–32 (2019) (arguing that process-based algorithmic accountability is prone to corporate co-optation and that migrating governance into code shifts power to engineers and their firms, and urging substantive audits/mandates).

[6] OpenAI, Response to the National Science Foundation [NSF] and Office of Science and Technology Policy [OSTP] Request for Information on the Development of an AI Action Plan 1 (Mar. 13, 2025), https://cdn.openai.com/global-affairs/ostp-rfi/ec680b75-d539-4653-b297-8bcf6e5f7686/openai-response-ostp-nsf-rfi-notice-request-for-information-on-the-development-of-an-artificial-intelligence-ai-action-plan.pdf [https://perma.cc/8EUH-7QT8] [hereinafter *OpenAI's Response*]; Google, Response to the NSF and OSTP Request for Information on the Development of an AI Action Plan 2 (Mar. 13, 2025), https://static.googleusercontent.com/media/publicpolicy.google/en//resources/response_us_ai_action_plan.pdf [https://perma.cc/SC8Q-HQSR] [hereinafter *Google's Response*].

While scholars have challenged the narrative of an AI arms race and its zero-sum logic,[7] this Article examines something more specific: how that narrative functions as an argument for deregulation. The "Regulation Sacrifice"[8]—the claim that nations must abandon safety oversight to win the AI race—has become a dominant policy paradigm, wielded by companies and governments alike to justify dismantling governance frameworks.[9] Rather than debating whether the arms race metaphor is apt or whether cooperation would be preferable, this Article accepts the Regulation Sacrifice on its own terms and asks: Does deregulation actually deliver the national security benefits it promises? Through systematic analysis across three time horizons—examining whether AI advantages prove durable, whether regulation actually impedes innovation, and whether deregulation enhances security—this Article demonstrates that sacrificing governance for speed fails not despite its national security rationale but because of it. The paradigm does not merely rest on a flawed metaphor; it fundamentally misconceives how security is achieved in an era of rapidly diffusing technology.

---

[7] *E.g.*, Peter Asaro, *What Is an 'Artificial Intelligence Arms Race' Anyway?*, 15 I/S: J.L. & POL'Y FOR INFO. SOC'Y 45 (2019); Paul Scharre, *Debunking the AI-Arms-Race Theory*, 4 TEX. NAT'L SEC. REV. 121 (2021); Stefka Schmid et al., *Arms Race or Innovation Race? Geopolitical AI Development*, 30 GEOPOLITICS 1907 (2025); Tiffany C. Li, *Ending the AI Race: Regulatory Collaboration as Critical Counter-Narrative*, 69 VILL. L. REV. 981 (2025); Kimberly A. Houser & Anjanette H. Raymond, *It Is Time to Move Beyond the 'AI Race' Narrative: Why Investment and International Cooperation Must Win the Day*, 18 NW. J. TECH. & INTELL. PROP. 129 (2021); Seán S. Ó hÉigeartaigh, *The Most Dangerous Fiction: The Rhetoric and Reality of the AI Race* 12–15 (May 28, 2025) (unpublished manuscript) (on file with SSRN), https://dx.doi.org/10.2139/ssrn.5278644. Even critics acknowledge that the perception of competition shapes policy. *See also* Huw Roberts et al., *Global AI Governance: Barriers and Pathways Forward*, 100 INT'L AFFS. 1275, 1279 (2024) (noting that cooperative action is "is influenced by factors like states' threat perceptions, levels of trust and alignment of interests").

[8] This Article introduces the term "Regulation Sacrifice" to describe the systematic abandonment of safety oversight justified by competitive imperatives. For related analyses of this phenomenon, *see* Markus Trengove & Emre Kazim, Dilemmas in AI Regulation: An Exposition of the Regulatory Trade-Offs Between Responsibility and Innovation 3–5 (Apr. 6, 2022) (unpublished manuscript) (on file with SSRN), https://dx.doi.org/10.2139/ssrn.4072436 (documenting trade-offs between innovation and responsibility). To be clear, the Regulation Sacrifice dynamic predates AI and extends beyond it. *See, e.g.*, Julie E. Cohen, *Between Truth and Power: The Legal Constructions of Informational Capitalism*, 132 HARV. L. REV. 991, 1023–28 (2019) (examining how platform companies deploy competitive necessity narratives to resist regulation); Ryan Calo, *Artificial Intelligence Policy: A Primer and Roadmap*, 51 U.C. DAVIS L. REV. 399, 425–30 (2017) (documenting early emergence of innovation-over-safety framing in AI policy).

[9] *Winning the AI Race*, *supra* note 1; Martin Coulter, *OpenAI May Leave the EU if Regulations Bite—CEO*, REUTERS (May 25, 2023), https://www.reuters.com/technology/openai-may-leave-eu-if-regulations-bite-ceo-2023-05-24/ [https://perma.cc/4ZA8-L78N]; Kate Crawford & Jason Schultz, *Big Data and Due Process: Toward a Framework to Redress Predictive Privacy Harms*, 55 B.C. L. REV. 93, 118–22 (2014) (documenting resistance to algorithmic accountability using efficiency arguments).

This Article establishes this argument in three steps. Part I traces how the arms race narrative has evolved from rhetorical device to concrete policy, generating a reflexive equation of regulation with strategic disadvantage. It examines how government officials and AI firms strategically wield national-security rhetoric to promote the Regulation Sacrifice narrative—the view that every month of extra speed is worth more to national security than any risk reduction achieved through oversight. By focusing on the United States and China, Part I demonstrates how this competitive logic constrains regulatory ambition domestically and relegates international coordination to toothless declarations of principle.[10]

Part II interrogates the three implicit premises that must hold for the Regulation Sacrifice to make strategic sense. First, it tests whether temporary bursts of deregulated speed create durable technological advantages, finding that AI capabilities diffuse so rapidly that performance gaps collapse within months, not years.[11] Second, it examines whether governance measures actually impose prohibitive costs on innovation, discovering that well-designed regulation often enhances rather than hinders technological progress.[12] Third, it asks whether racing ahead without safeguards truly enhances national security, a question whose answer depends on the specific threats AI poses across different time horizons. This final inquiry sets the stage for Part III's systematic examination of how deregulation affects security in practice.

---

[10] *See infra* Part I.

[11] *See also* NESTOR MASLEJ ET AL., STAN. HUM.-CENTERED AI, ARTIFICIAL INTELLIGENCE INDEX REPORT 2025 14–15 (2025) (documenting performance gap collapse); Mackenzie Ferguson, *DeepSeek's Breakthrough: A New Era for AI with Less Compute Power*, OPENTOOLS (Dec. 29, 2024), https://opentools.ai/news/deepseeks-breakthrough-a-new-era-for-ai-with-less-compute-power [https://perma.cc/U565-LVJC].

[12] Much of the research in on the relationship between regulation and innovation deals with environmental regulation. *See e.g.,* Michael E. Porter & Claas van der Linde, *Toward a New Conception of the Environment–Competitiveness Relationship*, 9 J. Econ. Persps. 97 (1995) (arguing that properly designed environmental regulation can induce 'innovation offsets' that partially or more than fully compensate compliance costs); Antoine Dechezleprêtre & Misato Sato, The Impacts of Environmental Regulations on Competitiveness, 11 Rev. Env't Econ. & Pol'y 183 (2017) (reviewing empirical literature and concluding that competitiveness effects are generally small and that regulation can induce innovation in cleaner technologies).

Part III employs a temporal framework recently proposed by Anthropic CEO Dario Amodei to assess whether deregulation attains its stated security objectives.[13] In the near term, the gravest threats are cybersecurity breaches and machine-scale misinformation, risks that only regulation can meaningfully suppress. Deregulation, by stripping away provenance checks, disclosure mandates, and liability safeguards, leaves those hazards free to proliferate.[14] Over the medium horizon, the retreat from regulation accelerates the erosion of safeguards that once confined bioweapons to a handful of state laboratories. Frontier models will be able to walk even modestly resourced groups through virulence optimization, DNA synthesis, and aerosol delivery. Dismantling of oversight infrastructure—sequence-order screening, model risk evaluations, liability regimes—lets those capabilities spread unchecked.[15] In the long term, competitive dynamics surrounding potential artificial general intelligence (AGI) create powerful incentives that systematically undermine safety,[16] with deregulation removing the tools required to keep transformative AI under meaningful human control.[17] At each horizon, the evidence

---

[13] *Oversight of A.I.: Principles for Regulation: Hearing Before the Subcomm. on Priv., Tech. & the L. of the S. Comm. on the Judiciary*, 118th Cong. 6–8 (Jul. 25, 2023) (written testimony of Dario Amodei, CEO, Anthropic) [hereinafter Amodei Testimony], https://www.judiciary.senate.gov/imo/media/doc/2023-07-26_-_testimony_-_amodei.pdf [https://perma.cc/HR3E-X4QL].

[14] *See* COAL. FOR CONTENT PROVENANCE & AUTHENTICITY [C2PA], C2PA TECHNICAL SPECIFICATION V1.3, 2 (2023), https://c2pa.org/specifications/specifications/1.3/specs/_attachments/C2PA_Specification.pdf [https://perma.cc/8GTK-2RED]; Danielle Citron & Robert Chesney, *Deepfakes and the New Disinformation War*, FOREIGN AFFS., Jan./Feb. 2019, at 147, 159–60.

[15] BILL DREXEL & CALEB WITHERS, CENTER FOR NEW AMERICAN SECURITY, AI AND THE EVOLUTION OF BIOLOGICAL NATIONAL SECURITY RISKS: CAPABILITIES, THRESHOLDS, AND INTERVENTIONS 3, 19 (2024); DAVID LUCKEY ET AL., MITIGATING RISKS AT THE INTERSECTION OF ARTIFICIAL INTELLIGENCE AND CHEMICAL AND BIOLOGICAL WEAPONS 34-36 (2025), https://www.rand.org/content/dam/rand/pubs/research_reports/RRA2900/RRA2990-1/RAND_RRA2990-1.pdf [https://perma.cc/GXN3-VKPW]; Jonas B. Sandbrink, Artificial Intelligence and Biological Misuse: Differentiating Risks of Language Models and Biological Design Tools 6 (Dec. 23, 2023) (unpublished manuscript) (on file with arXiv), https://arxiv.org/pdf/2306.13952 [https://perma.cc/GRU9-C2FB].

[16] Sandra Mayson, *Bias In, Bias Out*, 128 YALE L.J. 2218, 2256–60 (2019) (analyzing how competitive pressures lead to deployment of biased systems); Solon Barocas & Andrew D. Selbst, *Big Data's Disparate Impact*, 104 CALIF. L. REV. 671, 712–18 (2016) (documenting how speed imperatives conflict with discrimination prevention).

[17] NICK BOSTROM, SUPERINTELLIGENCE: PATHS, DANGERS, STRATEGIES 127-43 (2014); DANIEL KOKOTAJLO ET AL., AI 2027: A SCENARIO FOR SUPERHUMAN AI 9 (2025), https://ai-2027.com/ai-2027.pdf [https://perma.cc/8UGL-W3EU] (illustrating that leaked model weights

reveals that racing without rules amplifies rather than mitigates the security threats nations seek to avoid.

The evidence converges on a clear conclusion: the Regulation Sacrifice fails on its own terms. Rather than securing national advantage, it hands adversaries the same dangerous capabilities while creating cascading vulnerabilities that no nation can control. The paradigm persists not because it delivers security but because it satisfies powerful interests: companies seeking freedom from oversight, politicians seeking simple narratives, and a public primed to see competition everywhere.[18] Yet the stakes are too high for such convenient fictions. We stand at a juncture where governance choices will determine whether AI enhances human flourishing or amplifies existential risks.[19] By exposing the bankruptcy of this central paradigm, this Article clears space for approaches that recognize what the evidence makes plain: in an era of rapidly diffusing technology, security comes not from racing without rules but from channeling innovation through governance frameworks that enhance both safety and progress. The path forward begins with abandoning the dangerous mythology of the Regulation Sacrifice—the very thinking that leads inexorably to Mutually Assured Deregulation.

## I. THE REGULATION SACRIFICE

We start by examining how the framing of artificial intelligence development as a geopolitical arms race is deployed as an anti-regulatory argument, particularly in the United States and China, the two nations at the forefront of developing cutting edge

---

can eliminate safety advantages and that as racing accelerates, diplomatic arms-control windows shrink, driving governments toward coercive options); Yoshua Bengio, Implications of Artificial General Intelligence on National and International Security 12 (Oct. 16, 2024) (unpublished manuscript) (on file with Aspen Strategy Grp.), https://www.aspeninstitute.org/wp-content/uploads/2024/10/Bengio_National-and-International-Security-in-an-AGI-Context_Final.pdf [https://perma.cc/2QCL-5MDN].

[18] Ó hÉigeartaigh, *supra* note 7; Daniel Kahneman & Jonathan Renshon, *Hawkish Biases*, *in* AMERICAN FOREIGN POLICY AND THE POLITICS OF FEAR 79, 79–96 (A. Trevor Thrall & Jane K. Cramer eds., 2009) (documenting systematic biases toward competitive strategies).

[19] *Compare* John O. McGinnis, *Accelerating AI*, 104 NW. U. L. REV. 1253, 1268–72 (2010) (arguing government should support "friendly AI" research under light-touch governance), *with* Matthew U. Scherer, *Regulating Artificial Intelligence Systems: Risks, Challenges, Competencies, and Strategies*, 29 HARV. J.L. & TECH. 353, 378–82 (2016) (warning that unregulated AI development could lead to public risks and security vulnerabilities).

AI systems. It shows that government officials and AI firms strategically wield national-security rhetoric to create what I call a Regulation Sacrifice. This is the view that because the decisive objective is to outrun adversaries (usually China or the United States) to achieve frontier capabilities—and to lock in that lead before norms or treaties can erode it—states should minimize regulatory constraints that might slow domestic developers. Temporary privacy and security risks are framed as an acceptable price so long as the sprint delivers a durable edge that can later be leveraged to impose rules from a position of strength. In short: win first, tidy up later.

By focusing primarily on the United States, with comparative insights from China, I demonstrate how the arms race narrative functions not merely as description but as a powerful political tool that (i) constrains regulatory ambition domestically and (ii) relegates international initiatives to toothless statements of principle, narrowing the scope of possible governance approaches.

Since 2022, a competitive, geopolitical framework has overtaken discussions of AI governance, recasting technological development as a zero-sum contest between nations where security imperatives consistently trump regulatory concerns.[20] This "arms race frame" positions AI advancement primarily as a contest for technological supremacy and protection against foreign threats, rather than a shared challenge requiring coordinated oversight.[21] Especially in the United States, but also in China and elsewhere,[22] policymakers and industry leaders increasingly invoke this competitive framework to resist domestic regulation and minimize international constraints.[23]

---

[20] ASIA PAC. TASK FORCE, *supra* note 2.

[21] Li, *supra* note 7, at 983–87.

[22] Yasmeen Serhan, *How Israel Uses AI in Gaza—And What It Might Mean for the Future of Warfare*, TIME (Dec. 19, 2024), https://time.com/7202584/gaza-ukraine-ai-warfare/ [https://perma.cc/K7GC-CJ58]; Sarita Chaganti Singh & Nikunj Ohri, *India's Finance Ministry Asks Employees to Avoid AI Tools like ChatGPT, DeepSeek*, REUTERS (Feb. 5, 2025), https://www.reuters.com/technology/artificial-intelligence/indias-finance-ministry-asks-employees-avoid-ai-tools-like-chatgpt-deepseek-2025-02-05/ [https://perma.cc/L93K-6R24]; *Outline of the Act on Promotion of Research and Development, and Utilization of AI-Related Technology*, CABINET OFF., GOV'T OF JAPAN, https://www8.cao.go.jp/cstp/ai/ai_hou_gaiyou_en.pdf [https://perma.cc/2SA2-TTAE].

[23] *See* Elena Chachko, *National Security by Platform*, 25 STAN. TECH. L. REV. 55, 64 (2022) (noting that critics describe platforms' security/geopolitics reforms as "window dressing" intended "to deflect public criticism and appease regulators," and observing that security considerations can complicate regulatory responses, including antitrust pressure); Kristen E. Eichensehr, *Digital*

In the United States, the arms race narrative functions as a central argument for limiting regulation. Policymakers routinely invoke a geopolitical "AI race" against China to argue for light-touch regulation at home.[24] The framing is explicit: "[t]he way to beat China in the AI race is to outrace them in innovation, not saddle AI developers with European-style regulations," proclaimed Senator Ted Cruz, announcing a 2025 Senate hearing on "Winning the AI Race."[25] This competitive mindset casts regulation as a potential handicap to U.S. technological supremacy. Indeed, upon taking office in 2025, President Trump swiftly rescinded his predecessor's AI executive order, dismissing its safeguards as "onerous" requirements that would "threaten American technological leadership."[26] In its place, Trump directed an AI Action Plan meant to "sustain and enhance America's global AI dominance," explicitly tying deregulation to the strategic aim of outpacing China.[27]

National security concerns, and not only commercial interests, lie at the heart of this framing. Vice President Vance articulated this perspective clearly at the 2025 Artificial Intelligence Action Summit in Paris, warning about "hostile foreign adversaries" weaponizing AI and authoritarian regimes using it to "strengthen their military, intelligence, and surveillance capabilities."[28] This security-centered narrative positions AI not merely as an economic asset but as critical infrastructure in a geopolitical struggle. Vance further cautioned European allies against "partnering with such regimes," framing the tech race as a struggle between democratic and authoritarian models of governance.[29]

---

*Switzerlands*, 167 U. PA. L. REV. 665, 698–703 (2019) (showing that U.S. platforms may not align with U.S. national security aims when unregulated).

[24] *See, e.g.*, *Winning the AI Race*, *supra* note 1 ("The way to beat China in the AI race is to outrace them in innovation, not saddle AI developers with European-style regulations."); *Global Networks at Risk: Securing The Future of Communications Infrastructure: Hearing Before the Subcomm. on Commc'ns & Tech. of the H. Comm. on Energy and Com.*, 119th Cong. 1 (2025) (statement of Rep. Hudson), https://energycommerce.house.gov/posts/chairman-hudson-delivers-opening-statement-at-subcommittee-on-communications-and-technology-hearing-on-ai-and-communications-infrastructure [https://perma.cc/7MFM-EWAA] (urging a "light regulatory touch . . . [and that] AI must be given the same opportunity").

[25] *Winning the AI Race*, *supra* note 1.

[26] *Fact Sheet: President Donald J. Trump Takes Action to Enhance America's AI Leadership*, *supra* note 3.

[27] Exec. Order No. 14,179, *supra* note 3; THE WHITE HOUSE, WINNING THE RACE: AMERICA'S AI ACTION PLAN 3–6, 20–23 (2025), https://www.whitehouse.gov/wp-content/uploads/2025/07/Americas-AI-Action-Plan.pdf [https://perma.cc/J6VW-T35B].

[28] Vance, *supra* note 4.

[29] Vance, *supra* note 4.

Major AI companies' responses to the federal government's AI Action Plan Request for Information reveal how thoroughly industry has adopted this national security framing.[30] OpenAI warns of "a Chinese Communist Party (CCP) determined to overtake us by 2030" and argues for policies ensuring that "American-led AI . . . prevails over CCP-built autocratic, authoritarian AI."[31] OpenAI CEO Sam Altman has likewise framed AI development as a geopolitical race, warning that "The United States has got to lead in A.I. innovation. This technology is going to be used by our adversaries."[32] Similarly, Anthropic supports the Office of Science and Technology Policy's (OSTP) "effort to . . . maintain and strengthen America's dominance" while emphasizing the need for "comprehensive government awareness . . . as China advances its efforts to build powerful dual-use AI systems."[33] Google likewise frames AI development as a "global AI competition" where "policy decisions will determine the outcome."[34]

When policymakers conflate AI leadership with national security, they create a nearly irresistible argument against meaningful regulation. This security framing transforms what could be deliberative policy choices—with various legitimate resolutions—into binary questions of national survival, where regulatory leniency becomes the only patriotic option. In a regulatory proposal submission, OpenAI argued that state-level regulatory initiatives "could impose burdensome compliance requirements that may hinder [U.S.] economic competitiveness and undermine [U.S.] national security."[35] The company proposed "creating a tightly-scoped framework for voluntary partnership between the federal government and the private sector to expand and strengthen American national security" and argued that the "patchwork" of proposed AI regulations "risks bogging down innovation and, in the case of AI, undermining America's leadership position."[36] Google similarly called for "federal pre-emption of state-level laws that affect frontier AI models," arguing that this "would ensure a unified national framework . . . focused on protecting national security while fostering an environment where American AI innovation can thrive."[37]

---

[30] *See, e.g.*, *OpenAI's Response*, *supra* note 6, at 5 (urging an AI Action Plan whose goal is to "strengthen American national security."); *Google's Response*, *supra* note 6, at 1–2 ("Google welcomes the . . . goal to 'sustain and enhance America's global AI dominance' . . . . We are in a global AI competition . . . policy decisions will determine the outcome."); Anthropic, Response to the NSF and OSTP Request for Information on the Development of an Artificial Intelligence (AI) Action Plan 1–3 (Mar. 6, 2025), https://assets.anthropic.com/m/4e20a4ab6512e217/original/Anthropic-Response-to-OSTP-RFI-March-2025-Final-Submission-v3.pdf [https://perma.cc/5ZRB-43A7] [hereinafter *Anthropic's Response*] ("Our recommendations encompass . . . national security imperatives . . . . [A]rtificial intelligence systems . . . will present significant national security implications," so the administration "must strengthen national security frameworks . . . ."); Amazon, Response to the NSF and OSTP Request for Information on the Development of an Artificial Intelligence (AI) Action Plan 2 (2025), https://files.nitrd.gov/90-fr-9088/Amazon-AI-RFI-2025.pdf [https://perma.cc/7SNB-BYTT] ("Our economic and national security interests are best served by policies that enhance technological dynamism and open markets.").

[31] *OpenAI's Response*, *supra* note 6, at 1.

[32] MAKING MONEY: *FBN Interview with Sam Altman, CEO, OpenAI and Brian Schimpf, Co-Founder and CEO, Anduril Industries* (Fox Bus. Network television broadcast, aired Dec. 4, 2024), https://www.tradingview.com/news/reuters.com%2C2024-12-04%3Anewsml_CqtdgbYBa%3A0-fbn-interview-with-sam-altman-ceo-openai-and-brian-schimpf-co-founder-and-ceo-anduril-industries [https://perma.cc/RA2B-Z9KG].

[33] *Anthropic's Response, supra* note 30, at 2.

[34] *Google's Response*, *supra* note 6, at 1.

Such pleas for regulatory relief in exchange for voluntary cooperation reveal how directly companies link geopolitical rivalry to domestic policy choices. In its submission, Anthropic urged the government to "coordinate a cross-agency effort to identify and eliminate regulatory and procedural barriers to rapid AI deployment" and warned that "failing to address these energy requirements presents serious risks to America's technological leadership, as U.S. developers may be forced to relocate overseas."[38] Anthropic highlighted that "authoritarian regimes . . . are already actively courting American AI companies."[39] Both Anthropic and OpenAI emphasized the need for government-industry partnership on security issues, proposing that federal agencies develop "robust capabilities to rapidly assess any powerful AI system"[40] and provide "classified threat intelligence to mitigate national security risks" from frontier models and state actors.[41]

Beyond policy submissions, AI companies have increasingly positioned themselves as essential to national security interests. Google has shifted its stance to align more closely with U.S. security priorities, framing AI advancement as vital for national security and stating that "democracies should lead" in the "global competition."[42] This represents a notable pivot from Google's earlier reluctance to engage in military AI applications, with the company quietly dropping its 2018 pledge not to use AI for weapons systems.[43] Similarly, at a Senate hearing, Microsoft president Brad Smith cautioned that preserving U.S. leadership requires letting "academics and entrepreneurs . . . innovate and deploy models without huge barriers."[44] Facebook founder Mark Zuckerberg prepared talking points for his 2018 Senate testimony describing "US tech companies [as a] key asset for America" and warning that breaking up Facebook would "strengthen[] Chinese companies," casting Big Tech as the nation's champion in its competition with China.[45]

---

[35] *OpenAI's Response*, *supra* note 6, at 4.
[36] *Id.* at 6.
[37] *Google's Response*, *supra* note 6, at 5.
[38] *Anthropic's Response*, *supra* note 30, at 7-8.
[39] *Id.* at 7.
[40] *Id.* at 3.
[41] *OpenAI's Response*, *supra* note 6, at 6.

Anthropic's leadership has similarly framed AI safety in competitive terms. CEO Dario Amodei warned Congress of "extraordinarily grave threats to U.S. national security over the next 2 to 3 years" from AI, while arguing that pausing U.S. AI development would only "hand over its power . . . to adversaries who do not share our values."[46] This characterization, revisited below, presents AI development as a race to manage severe risks before adversaries can exploit them: a framing that encourages acceleration rather than regulation.[47]

Lawmakers have echoed this rationale in congressional debates on technology funding. In arguing for the CHIPS and Science Act and similar measures, members of Congress emphasized that heavy-handed rules could undercut the U.S. innovation edge vis-à-vis

---

[42] James Manyika & Demis Hassabis, *Responsible AI: Our 2024 Report and Ongoing Work,* GOOGLE BLOG (Feb. 4, 2025), https://blog.google/technology/ai/responsible-ai-2024-report-ongoing-work/ [https://perma.cc/9VKH-MTNH].
[43] *Id. See also,*
[44] Faiza Patel & Melanie Geller, *Senate AI Hearings Highlight Increased Need for Regulation*, BRENNAN CTR. FOR JUST. (Oct. 13, 2023), https://www.brennancenter.org/our-work/analysis-opinion/senate-ai-hearings-highlight-increased-need-regulation [https://perma.cc/ZK93-EZJW].
[45] Nick Statt, *Read Mark Zuckerberg's Notes from Today's Facebook Privacy Senate Hearing,* THE VERGE (Apr. 10, 2018), https://www.theverge.com/2018/4/10/17222546/facebook-mark-zuckerberg-senate-hearing-notes-cambridge-analytica-privacy [https://perma.cc/25NV-HVGA].
[46] Amodei Testimony, *supra* note 13, at 1.
[47] Sam Meacham, *A Race to Extinction: How Great Power Competition Is Making Artificial Intelligence Existentially Dangerous*, HARV. INT'L REV. (Sept. 8, 2023) https://hir.harvard.edu/a-race-to-extinction-how-great-power-competition-is-making-artificial-intelligence-existentially-dangerous/ [https://perma.cc/52XQ-DVVY].

China.[48] As one House Republican put it bluntly during a recent hearing, "we are in an arms race with China for artificial intelligence."[49] This arms race framing has fostered a deregulation sacrifice in domestic policy. Both parties agree that AI policy must prioritize innovation and prevent China from dominating the field, with Republicans warning against regulatory overreach and Democrats emphasizing that well-designed guardrails can promote rather than stifle U.S. leadership.[50] At a 2023 Senate hearing, the tech industry's leading trade group warned that excessive regulation would "undermine" U.S. AI leadership and simply hand an advantage to "authoritarian nations."[51]

From the government and civil society to Republicans and Democrats, then, the prevailing sentiment is that the United States can't afford to lose the AI race—so any regulation must be carefully calibrated not to hamper the nation's ability to outcompete China.[52] The result is a reluctance to impose strict domestic constraints absent comparable commitments by adversaries. In short, fear of falling behind has become a powerful political argument against AI regulation.

---

[48] Khushboo Razdan, *US-China Tech War: Second-Tier 'Legacy Chips' at Forefront of Battle for Semiconductor Supremacy*, S. CHINA MORNING POST (Feb. 23, 2024), https://www.scmp.com/news/china/article/3253031/us-china-tech-war-second-tier-legacy-chips-forefront-battle-semiconductor-supremacy [https://perma.cc/ZY4S-CZDP].

[49] *America's AI Moonshot: The Economics of AI, Data Centers, and Power Consumption: Hearing Before the H. Comm. on Oversight & Gov't Reform*, 119th Cong. 15 (2025) (statement of Rep. Gary Palmer, Member, H. Subcomm. on Econ. Growth, Energy Pol'y & Regul. Affs.), https://www.congress.gov/event/119th-congress/house-event/118078/text [https://perma.cc/7GYV-ACZL].

[50] *Winning the AI Race*, *supra* note 1 (Senator Cruz claimed, "The way to beat China in the AI race is to outrace them in innovation, not saddle AI developers with European-style regulations."); *Majority Leader Schumer Delivers Remarks to Launch SAFE Innovation Framework for Artificial Intelligence At CSIS*, SENATE DEMOCRATS (June 21, 2023), https://www.democrats.senate.gov/news/press-releases/majority-leader-schumer-delivers-remarks-to-launch-safe-innovation-framework-for-artificial-intelligence-at-csis [https://perma.cc/2F4Y-SFS6] ("Innovation must be our north star . . . encouraging, not stifling, innovation," and warning that the Chinese Communist Party "could leap ahead . . . and set the rules of the game for AI.").

[51] Patel & Geller, *supra* note 44.

[52] *See e.g., Remarks by President Biden in a Farewell Address to the Nation*, THE WHITE HOUSE (Jan. 15, 2025), https://bidenwhitehouse.archives.gov/briefing-room/speeches-remarks/2025/01/15/remarks-by-president-biden-in-a-farewell-address-to-the-nation/ [https://perma.cc/A7ZG-BCQ3] (warning that "unless safeguards are in place, AI could spawn new threats," while insisting that "America — not China — must lead the world on the development of AI"); Cristiano Lima-Strong, *Transcript: Sam Altman Testifies at US Senate Hearing on AI Competitiveness*, TECH POL'Y PRESS (May 8, 2025), https://www.techpolicy.press/transcript-sam-altman-testifies-at-us-senate-hearing-on-ai-competitiveness/ [https://perma.cc/8XTY-SQ4C] (reflecting bipartisan "AI race" framing—e.g., "America has to beat China in the AI race" (Cruz) and "we want to win against China" (Cantwell)—while urging "sensible regulation that does not slow us down" (Altman)).

A comparable, *though not identical*, "regulation-sacrifice" dynamic operates in Beijing.[53] Party leaders frame artificial intelligence as a zero-sum contest for national survival and then pare back any restraint that might slow deployment. Presiding over the Politburo's twentieth collective study session on April 25, 2025, President Xi Jinping called AI "a strategic technology that is leading the new round of scientific and technological revolution," urged cadres to "master[] core technologies such as high-end microchips and foundational software," and warned that the Party must prepare for the technology's "unprecedented risks and challenges."[54] His battlefield rhetoric is not new. At an earlier study session on October 31, 2018, he ordered China to "occup[y] the high ground in critical and core AI technologies."[55]

Other senior figures echo the same theme. Addressing the U.N. Security Council's debate on AI, Ambassador Zhang Jun cast the field in explicitly geopolitical terms, accusing "a certain developed country" of pursuing "technological hegemony," "build[ing] exclusive small clubs," "maliciously obstruct[ing] the technological development of other countries," and "artificially creat[ing] technological barriers." He paired that competition framing with a military warning—stressing that "AI in the military field may lead to major changes in the way of warfare and the format of war," and urging states to "oppose the use of AI to seek military hegemony." [56] Reinforcing the point, Tsinghua University computer scientist Wen Gao wrote in *People's Daily* that AI is the "strategic commanding height for competition among major powers."[57]

---

[53] Because internal policy deliberations in Beijing remain opaque, scholars must infer intent from public messaging and successive rule revisions rather than from open hearings or comment dockets—making the evidentiary record necessarily more circumstantial than in the United States.

[54] Xi Jinping Zai Zhonggong Zhongyang Zhengzhi Ju Di Ershi Ci Jiti Xuexi Shi Qiangdiao Jianchi Zili Ziqiang Tuchu Yingyong Daoxiang Tuidong Rengong Zhineng Jiankang You Xu Fazhan (习近平在中共中央政治局第二十次集体学习时强调 坚持自立自强 突出应用导向 推动人工智能健康有序发 展), XINHUA NEWS AGENCY (Apr. 26, 2025), https://www.news.cn/politics/leaders/20250426/63f5cde8f4b54e22ba35aa7ec7884b3a/c.html [https://perma.cc/5UBL-GCVG] *translated in At the 20th Collective Study Session of the CCP Central Committee Politburo, Xi Jinping Stresses: Persist in Being Self-Reliant, Be Strongly Oriented Toward Applications, and Push the Orderly Development of Artificial Intelligence*, CTR. FOR SEC. & EMERGING TECH. (Apr. 28, 2025), https://cset.georgetown.edu/publication/xi-politburo-collective-study-ai-2025/ [https://perma.cc/XTQ8-QQJ3].

[55] Xi Jinping (习近平), Xi Jinping: Tuidong Woguo Xin Yidai Rengong Zhineng Jiankang Fazhan (习近平：推动我国新一代人工智能健康发展), XINHUA NEWS AGENCY (Oct. 31, 2018), http://www.xinhuanet.com/politics/2018-10/31/c_1123643321.htm [https://perma.cc/T595-J24Q] *translated in* Rogier Creemers & Elsa B. Kania, *Translation: Xi Jinping Calls for 'Healthy Development' of AI* , DIGICHINA, (Nov. 5, 2018), https://digichina.stanford.edu/work/xi-jinping-calls-for-healthy-development-of-ai-translation/ [https://perma.cc/YB5E-LX44].

Scholars suggest that such security-race framing shapes Chinese regulation. The final July 2023 *Interim Measures for Generative-AI Services* retained strict content rules but scrapped the draft's blanket ex-ante licensing, changes Reuters judged "far less onerous" because Beijing "wants to support generative-AI tech development."[58] Crucially, the Measures still mandate algorithm filings, security assessments, and adherence to "core socialist values"; the only substantive easings were the removal of blanket licensing and an unrealistic truth-accuracy clause.[59] Professor Angela Huyue Zhang likewise argues that China's "strategic[ally] lenient approach to regulation may therefore offer its AI firms a short-term competitive advantage over their European and U.S. counterparts," using permissive AI rules to give domestic firms an edge in the global race.[60]

---

[56] *Remarks by Ambassador Zhang Jun at the U.N. Security Council Briefing on Artificial Intelligence: Opportunities and Risks for International Peace and Security*, PERMANENT MISSION OF THE PEOPLE'S REPUBLIC OF CHINA TO THE U.N. (July 18, 2023), https://un.china-mission.gov.cn/eng/chinaandun/securitycouncil/202307/t20230719_11114947.htm.

[57] Gao Wen (高文), *Qiangzhua Rengong, Zhineng Fazhan de Lishixing Jiyu* (抢抓人工智能发展历史性机遇) [*Seizing the Historic Opportunity for the Development of Artificial Intelligence*], ZHONGGUO GONGCHANDANG XINWEN WANG (中国共产党新闻网) [CPCNEWS] (Feb. 24, 2025), https://cpc.people.com.cn/n1/2025/0224/c64387-40424141.html [https://perma.cc/P9KB-V7VT].

[58] Josh Ye, *China Says Generative AI Rules to Apply Only to Products for the Public*, REUTERS (July 13, 2023), https://www.reuters.com/technology/china-issues-temporary-rules-generative-ai-services-2023-07-13/ [https://perma.cc/6LMY-LRQG].

[59] Shengchengshi Rengong Zhineng Fuwu Guanli Zhangxing Banfa (生成式人工智能服务管理暂行办法) [Interim Measures for the Management of Generative Artificial Intelligence Services] Chapter 3 (promulgated by the Cyberspace Admin. of China, July 10, 2023, effective Aug. 15, 2023) St. Council Gaz., Aug. 30, 2023, at 40, https://www.gov.cn/gongbao/2023/issue_10666/ [https://perma.cc/B6DK-GVBA] (China).

[60] Angela Huyue Zhang, *The Promise and Perils of China's Regulation of Artificial Intelligence*, 63 COLUM. J. TRANSNAT'L L. 1, 1 (2025).

Abroad, Beijing markets the same stance of "responsible AI": at the 2023 U.K. AI-Safety Summit, Wu Zhaohui, Vice President of the Chinese Academy of Sciences, insisted that every country enjoys an "equal right to develop and use AI," implicitly warning against restrictive regimes. [61] In short, China publicly champions responsible innovation while ensuring that oversight at home remains flexible enough to keep the nation sprinting for the AI commanding heights. While U.S. debates revolve around preserving private-sector agility, China's adjustments are made within a one-party system that still prioritizes ideological control. Thus, its deregulatory moves are narrower and principally aimed at shielding development speed from new procedural hurdles rather than relaxing existing content and security mandates.[62]

The arms race rationale also undercuts momentum for robust international coordination. A race mindset makes each major power wary of binding limits, lest rules restrain it more than its rival. Washington rejects proposals for a global AI "pause" or strict multilateral caps on frontier models, arguing that authoritarian competitors would simply ignore them; U.S. officials instead back export controls on chips and voluntary pacts among allies.[63] Beijing reciprocates. While its Position Paper on Global AI Governance endorses general principles of responsible development, it warns against any framework that might contain China's AI progress, signaling a red line against enforceable ceilings.[64]

Consequently, multilateral initiatives remain largely procedural and nonbinding. The 2023 Bletchley Declaration on AI Safety extols collective efforts to manage frontier risks yet makes only one binding commitment: to meet again.[65] The G-7 Hiroshima Process adopts the same logic, as its Advanced AI Model Risk Reports are voluntary so that no country inhibits its own AI competitiveness.[66] Both documents exemplify how great power rivalry converts ambitious governance proposals into minimalist pledges. As a result, Washington and Beijing engage in high-profile dialogue while quietly prioritizing national strategies and seeking to outdo each other. International coordination thus suffers the same regulation-sacrifice dynamic that shapes domestic oversight: everyone professes support for "responsible AI," but the arms race frame ensures that concrete constraints are deferred, diluted, or displaced.

---

[61] *China Willing to Enhance Communication with All Sides on AI Regulation—Vice Minister*, REUTERS (Nov. 1, 2023), https://www.reuters.com/technology/china-willing-enhance-communication-with-all-sides-ai-regulation-vice-minister-2023-11-01/ [https://perma.cc/X8GV-N6QS] (Wu stated that "[c]ountries regardless of their size and scale have equal rights to develop and use AI" and called for "global cooperation" and open-source availability).

[62] *See supra* notes 58–59 and accompanying text.

[63] Press Release, Bureau of Indus. & Sec, Commerce Strengthens Export Controls to Restrict China's Capability to Produce Advanced Semiconductors for Military Applications (Dec. 2, 2024) https://www.bis.gov/press-release/commerce-strengthens-export-controls-restrict-chinas-capability-produce-advanced-semiconductors-military [https://perma.cc/PNX9-GPH2].

[64] *Global AI Governance Initiative*, MINISTRY OF FOREIGN AFFS. OF THE PEOPLE'S REPUBLIC OF CHINA (Oct. 20, 2023), https://www.mfa.gov.cn/eng/zy/gb/202405/t20240531_11367503.html [https://perma.cc/D894-C7SY].

In short, the pervasive framing of AI as a zero-sum race with national security implications has bred a policy impasse. Every major player professes support for "responsible AI" globally, but meaningful coordination falters as each strives for the lead.[67] The arms race frame thus drives a cycle of competitive deregulation at home and cautious, weak engagement abroad—a dynamic termed here the Regulation Sacrifice. This prompts us to consider whether racing ahead in AI unchecked is truly a sound strategic choice or a dangerous policy trap. This raises the question which this Article seeks to answer: is sacrificing regulation for a victory in the "AI arms race" actually likely to lead to significant national security gains?

---

[65] *The Bletchley Declaration by Countries Attending the AI Safety Summit, 1-2 November 2023*, GOV.UK (Feb. 13, 2025) [hereinafter *The Bletchley Declaration*], https://www.gov.uk/government/publications/ai-safety-summit-2023-the-bletchley-declaration/the-bletchley-declaration-by-countries-attending-the-ai-safety-summit-1-2-november-2023 [https://perma.cc/G7A7-943Y].

[66] *OECD Launches Global Framework to Monitor Application of G7 Hiroshima AI Code of Conduct,* ORG. FOR ECON. CO-OP. & DEV. (Feb. 7, 2025), https://www.oecd.org/en/about/news/press-releases/2025/02/oecd-launches-global-framework-to-monitor-application-of-g7-hiroshima-ai-code-of-conduct.html [https://perma.cc/9RL6-KWD6].

[67] *Compare The Bletchley Declaration*, *supra* note 65, *with Fact Sheet: President Biden Issues Executive Order on Safe, Secure, and Trustworthy Artificial Intelligence*, THE WHITE HOUSE (Oct. 30, 2023), https://bidenwhitehouse.archives.gov/briefing-room/statements-releases/2023/10/30/fact-sheet-president-biden-issues-executive-order-on-safe-secure-and-trustworthy-artificial-intelligence/ [https://perma.cc/7CGN-CVXW], *and Global AI Governance Initiative*, *supra* note 64.

## II. THE ASSUMPTIONS OF THE AI REGULATION SACRIFICE

Part I traced how the arms race frame has generated a policy reflex that treats regulation as a strategic liability. The distilled doctrine, which this Article calls the Regulation Sacrifice, can be stated succinctly: In a geopolitical sprint for frontier AI, every month of additional speed is worth more to national security than any risk reduction achieved through near-term oversight. Therefore, states should defer or dilute regulation until after they have secured an unassailable lead. In effect, policymakers wager that they can win first and tidy up later.

For that wager to be plausible, three implicit premises must all hold:

1. *Durable Lead Assumption.* A temporary burst of deregulated speed can be turned into a long-lasting technological gap that rivals cannot quickly erase or leap-frog.
2. *Low Drag Assumption.* Governance measures, such as licensing, red-teaming mandates, and disclosure rules, impose costs or delays large enough to matter in the race; stripping them away yields a meaningful acceleration dividend.
3. *Net Security Benefit Assumption.* Achieving the frontier first confers security advantages that exceed the new vulnerabilities created by rapid, lightly governed deployment.

Let us interrogate each premise in turn, asking whether the empirical record and technical realities justify the confidence the Regulation Sacrifice requires.

### *A. Durable Lead Assumption*

The Regulation Sacrifice depends on a foundational premise: that early advantages in AI development can be converted into a lasting geopolitical edge. This assumption permeates current policy discourse. The Trump Administration's Executive Order 14179 exemplifies this thinking, pledging to "sustain and enhance America's global AI dominance."[68] Google's submission to the OSTP reinforces this narrative, asserting that "we are in a global AI competition, and policy decisions will determine the outcome. A pro-innovation approach that protects national security and ensures that everyone benefits from AI is essential to realizing AI's transformative potential and ensuring that America's lead endures."[69]

[68] Exec. Order No. 14,179, *supra* note 3, § 1.

Proponents of rapid, unregulated development typically invoke durability mechanisms. First, they point to scale advantages and capital barriers—early leaders capture infrastructure investments that competitors struggle to replicate,[70] with frontier models requiring multibillion-dollar data centers.[71] Second, they emphasize hardware-based technological gaps maintained through export controls on advanced semiconductors.[72] Third, they cite platform integration and the costs of migrating AI models embedded in existing platforms to new ones.[73]

These mechanisms share a common thread: they assume temporary speed advantages translate into permanent strategic positions. Without this durability, the logic of sacrificing safety for speed collapses. If advantages prove ephemeral, then we must ask whether a technological head start of even a year or two justifies abandoning protective oversight—a question whose answer depends critically on whether racing enhances security, as Part III will examine below.

Recent data reveals a stark reality: AI advantages seem to erode far more rapidly than the arms race narrative acknowledges. The most compelling evidence comes from direct measurement of capability gaps. Stanford's 2025 AI Index reports rapid U.S.–China convergence at the frontier: on the LMSYS Chatbot Arena, the top U.S. model's lead over the best Chinese model shrank from 9.3% in January 2024 to 1.7% by February 2025.[74] On specific benchmarks, On standard benchmarks, gaps also narrowed sharply from end-2023 to end-2024—for example, MMLU fell from 17.5 to 0.3 percentage points and MATH from 24.3 to 1.6 percentage points.[75]

---

[69] *Google's Response*, *supra* note 6.

[70] *Accord* Julie E. Cohen, *Law for the Platform Economy*, 51 U.C. DAVIS L. REV. 133, 156–61 (2017) (explaining how dominant platforms leverage protocol-based control, preferential placement, and interplatform affiliation to make participation 'sticky' and translate scale into durable competitive advantage ).

[71] Sebastian Moss, *Here's the 5GW, 30 Million Sq Ft Data Center Pitch Doc OpenAI Showed the White House*, DATA CENTER DYNAMICS (Nov. 6, 2024), https://www.datacenterdynamics.com/en/news/heres-the-5gw-30-million-sq-ft-data-center-pitch-doc-openai-showed-the-white-house/ (reporting OpenAI's internal analysis that a 5GW AI data center "could cost $100 billion").

[72] *See generally* CHRIS MILLER, CHIP WAR: THE FIGHT FOR THE WORLD'S MOST CRITICAL TECHNOLOGY (2022) (describing U.S. efforts to maintain semiconductor dominance through export controls).

[73] Colton Stradling, *Office365 Tops 400 Million Paid Users. Microsoft Credits Copilot AI for Office Growth*, YAHOO!TECH (Jan. 31, 2024), https://tech.yahoo.com/ai/articles/office365-tops-400-million-paid-204853643.html [https://perma.cc/AHH6-MPMH] (reporting over 400 million Office 365 seats); Lina M. Khan, *The Separation of Platforms and Commerce*, 119 COLUM. L. REV. 973, 1012 (2019) (documenting how dominant platforms' discriminatory conduct and "self-privileging" can "corrupt[] the entire system of platform-based innovation," ).

This rapid convergence occurs despite substantial U.S. advantages in both compute capacity and investment. The United States possesses approximately ten times more AI computing infrastructure than China and attracted $109 billion in private AI investment in 2024 compared to China's $9.3 billion.[76] Yet these resource advantages have not translated into durable capability gaps. The reason lies in the unique characteristics of AI technology that accelerate diffusion through multiple channels.

First, and perhaps most important, is the dramatic improvement in computational efficiency that has sharply accelerated AI development. Inference costs fell 280-fold between November 2022 and October 2024, from $20 to $0.07 per million tokens.[77] Training costs show similar declines. Achieving GPT-3 level performance cost $4.6 million in 2020 but only $450,000 by 2022, a 70% annual reduction.[78]

These efficiency gains enable competitors to achieve frontier capabilities with dramatically fewer resources. DeepSeek's January 2025 breakthrough exemplifies this dynamic: the Chinese firm achieved performance comparable to OpenAI's o1 model using only $6 million in training costs versus the roughly $100 million spent by U.S. competitors: an improvement in efficiency by a factor of fifteen to twenty,[79] although the overall development cost was likely much higher.[80] Critically, DeepSeek accomplished this using

---

[74] MASLEJ ET AL., *supra* note 11, at 97, fig. 2.1.36 and accompanying text.
[75] MASLEJ ET AL., *supra* note 11, at 97, fig. 2.1.37.
[76] Lennart Heim, *China's AI Models Are Closing the Gap—but America's Real Advantage Lies Elsewhere*, RAND (May 2, 2025), https://www.rand.org/pubs/commentary/2025/05/chinas-ai-models-are-closing-the-gap-but-americas-real.html; MASLEJ ET AL., *supra* note 11 at 3.
[77] *Id.* at 64.
[78] Haziqa Sajid, *AI Training Costs Continue to Plummet*, UNITE.AI (Mar. 16, 2023), https://www.unite.ai/ai-training-costs-continue-to-plummet/.
[79] Ferguson, *supra* note 11.
[80] Charles Mok, *Taking Stock of the DeepSeek Shock*, STAN. CYBER POL'Y CTR. (Feb. 5, 2025), https://cyber.fsi.stanford.edu/publication/taking-stock-deepseek-shock [https://perma.cc/38JX-5AUN].

export-restricted H800 chips, demonstrating how algorithmic innovation can overcome hardware limitations.[81]

Second, the AI research ecosystem operates with unprecedented openness that facilitates rapid knowledge transfer. In 2023, 65.7% of foundation models were released as open-source, up from 44.4% in 2022.[82] Meta's LLaMA model leak in March 2023 sparked an entire ecosystem of derivative models, demonstrating how quickly controlled AI systems can escape containment.[83] Even proprietary systems face security challenges—OpenAI experienced a significant breach in 2023 that exposed internal AI development details, while multiple ChatGPT incidents affected hundreds of thousands of users.[84]

Beyond intentional releases and breaches, knowledge flows through human capital mobility. Research shows that as of 2022, forty-seven percent of top AI researchers had Chinese undergraduate backgrounds, with high turnover rates facilitating rapid knowledge transfer between organizations and countries.[85] While NDAs, trade secret law, and other employment contracts may limit the disclosure of specific information, they do not prevent the diffusion of tacit expertise, methods, and generalizable insights as researchers change jobs in a highly competitive global market.[86]

---

[81] *Id.*

[82] NESTOR MASLEJ ET AL., STAN. HUM.-CENTERED AI, ARTIFICIAL INTELLIGENCE INDEX REPORT 2025 57, fig. 1.3.14 (2024).

[83] James Vincent, *Meta's Powerful AI Language Model Has Leaked Online—What Happens Now?*, THE VERGE (Mar. 8, 2023), https://www.theverge.com/2023/3/8/23629362/meta-ai-language-model-llama-leak-online-misuse [https://perma.cc/W9LG-M2SS].

[84] *See, e.g.*, *OpenAI's Internal AI Details Stolen in 2023 Breach*, REUTERS (July 5, 2024), https://www.reuters.com/technology/cybersecurity/openais-internal-ai-details-stolen-2023-breach-nyt-reports-2024-07-05/ [https://perma.cc/9T8A-BE6R]; *Group-IB Discovers 100K+ Compromised ChatGPT Accounts on Dark Web Marketplaces; Asia-Pacific Region Tops the List*, GROUP-IB (June 20, 2023), https://www.group-ib.com/media-center/press-releases/stealers-chatgpt-credentials/ [https://perma.cc/U9RU-A5ZR] (finding 101,134 credentials in stealer logs); Ravie Lakshmanan, *Over 225,000 Compromised ChatGPT Credentials Up for Sale on Dark Web Markets*, THE HACKER NEWS (Mar. 5, 2024), https://thehackernews.com/2024/03/over-225000-compromised-chatgpt.html [https://perma.cc/Z9S8-G96K] (noting that over 225,000 credentials offered for sale by October 2023).

[85] *The Global AI Talent Tracker 2.0*, MACROPOLO, https://macropolo.org/digital-projects/the-global-ai-talent-tracker [https://perma.cc/7HC8-EZEL] (archived Dec. 1, 2025); *AI Skills Gap Widens*, RANDSTAD (Nov. 12, 2024), https://www.randstad.com/press/2024/ai-skills-gap-widens/ [https://perma.cc/TY3Q-J3RH] (reporting that AI-skilled workers with one to five years' experience had a thirty-three percent job change rate in the prior year)

[86] *See* RESTATEMENT (THIRD) OF UNFAIR COMPETITION § 41 (AM. L. INST. 1995) (noting that trade secret law imposes a duty of confidence on employees and other parties who receive information pursuant to a confidential relationship or contractual obligation, thereby

This talent circulation ensures that breakthrough insights rarely remain confined to single institutions or nations.

Third, the empirical record on semiconductor export controls reveals their limited effectiveness in maintaining durable advantages. While controls imposed immediate costs—Nvidia faced $8 billion in quarterly sales losses—they also triggered adaptive responses that may accelerate competition.[87] Chinese firms responded by stockpiling chips, finding cloud workarounds, and most significantly, accelerating domestic innovation.[88]

A comprehensive study of thirty semiconductor companies found no evidence that export controls hindered innovation among affected firms. Counter-intuitively, impacted companies showed higher research spending and patent filings, suggesting that controls strengthened rather than weakened competitive capabilities.[89] Hardware constraints also forced Chinese developers to pioneer efficiency-driven approaches that achieve comparable

---

restricting unauthorized disclosure); *See also* John Villasenor, *Artificial Intelligence, Trade Secrecy, and the Challenge of Transparency,* 25 N.C. J.L. & Tech. 495, 511–12 (2024) (explaining that trade secret law does not extend to employees' general knowledge, skill, and experience—i.e., know-how that can travel with workers as they move between employers); Camilla A. Hrdy & Christopher B. Seaman, *Beyond Trade Secrecy: Confidentiality Agreements that Act Like Noncompetes,* 133 Yale L.J. 669, 694–99 (2024) (arguing that confidentiality agreements become problematic when they purport to reach beyond legitimately secret information and thereby function like noncompetes by chilling mobility and the use of general skills).

[87] Jack Whitney, Matthew Schleich & William Alan Reinsch, *The Double-Edged Sword of Semiconductor Export Controls* 2–3, 7–8 (Ctr. for Strategic & Int'l Stud., Oct. 2024) (explaining that Chinese "design-out" and "design-around" responses—alongside circumvention—threaten to make export controls "less effective as a longer-term barrier" and can weaken U.S. firms by reducing market access and revenue that funds R&D); Tova Donovan Levin, Sydney Hsu & Karen Hyunbee Cho, *U.S. Sensitive Technology Export Controls Toward China: Lessons from the Soviet Union* 1–2 (Hoover Hist. Lab, Apr. 2023) (summarizing historical studies finding no "substantial evidence" that Cold War export controls produced significant impacts on Soviet technological development and cautioning that broad controls may be counterproductive or of uncertain efficacy); Arsheeya Bajwa & Stephen Nellis, *Nvidia Shares Rise as Sales Hit from China Export Curbs Not as Bad as Feared*, REUTERS (May 29, 2025), https://www.reuters.com/business/nvidia-forecasts-second-quarter-revenue-below-estimates-2025-05-28/ (reporting that new U.S. export restrictions would "slice off $8 billion in sales" from Nvidia's then-current quarter).

[88] Gregory C. Allen, *China's New Strategy for Waging the Microchip Tech War*, CTR. FOR STRATEGIC & INT'L STUD. (May 3, 2023), https://www.csis.org/analysis/chinas-new-strategy-waging-microchip-tech-war#:~:text=In%20the%20private%20sector%2C%20Chinese,hedge%20against%20potential%20future%20restrictions [https://perma.cc/P6B2-AHG9].

[89] Andreas Schumacher, *Did U.S. Semiconductor Export Controls Harm Innovation?*, CTR. FOR STRATEGIC & INT'L STUD. (Nov. 5, 2024), https://www.csis.org/analysis/did-us-semiconductor-export-controls-harm-innovation [https://perma.cc/QNY7-8AC3] (finding no negative innovation impact across thirty firms).

results with dramatically fewer resources, creating competitive threats that may prove more difficult to counter than straightforward hardware superiority.[90]

The evidence converges on a clear conclusion: in AI, first-mover advantages will likely not last very long before competitors achieve functional parity.[91] This finding raises the critical question at the heart of the Regulation Sacrifice—is a short-term technological lead worth abandoning safety oversight?

The answer depends entirely on what that temporary advantage delivers for national security. If a year of AI superiority provided decisive strategic benefits—the ability to secure critical infrastructure, establish dominant standards, or achieve breakthroughs in defense capabilities—then the sacrifice might prove rational. But if racing merely hands adversaries the same dangerous capabilities months later while creating uncontrolled risks in the interim, then the bargain makes little strategic sense.

This temporal reality—advantages measured in months, not decades—fundamentally alters the cost-benefit calculation of the Regulation Sacrifice. Actors are not trading safety for long-term advantage, but rather permanent risks for temporary benefits. Whether even those temporary benefits enhance security remains to be seen. As Part III demonstrates, the evidence suggests that racing without safeguards systematically undermines rather than enhances national security across every time horizon.

---

[90] *See* DeepSeek-AI, DeepSeek-V2: A Strong, Economical, and Efficient Mixture-of-Experts Language Model 1 (June 19, 2024) (unpublished manuscript) (on file with arXiv), https://arxiv.org/pdf/2405.04434 [https://perma.cc/48SZ-GU9Z]; DeepSeek-AI, DeepSeek-V3 Technical Report 1 (Feb. 18, 2025) (unpublished manuscript) (on file with arXiv), https://arxiv.org/pdf/2412.19437 [https://perma.cc/H8LK-SZZJ]; John Villasenor, *DeepSeek Shows the Limits of U.S. Export Controls on AI Chips*, BROOKINGS (Jan. 29, 2025), https://www.brookings.edu/articles/deepseek-shows-the-limits-of-us-export-controls-on-ai-chips/ [https://perma.cc/8L29-QQLD] (arguing that controls "may actually be accelerating" Chinese progress by forcing engineers to learn to use limited compute more efficiently).

[91] Leonard Dung & Max Hellrigel-Holderbaum, *Against Racing to AGI: Cooperation, Deterrence, and Catastrophic Risks* 15 (arXiv, Working Paper No. 2507.21839, 2025) ("[P]ushing the frontier forward is more difficult than catching up to the frontier . . . ."); *but see* Janet Egan, Paul Scharre & Vivek Chilukuri, *Promote and Protect America's AI Advantage*, CTR. NEW AM. SEC. (Jan. 20, 2025), https://www.cnas.org/publications/commentary/promote-and-protect-americas-ai-advantage [https://perma.cc/77S8-XPD3] (arguing that "ensuring these breakthrough capabilities are developed in the United States is critical for both U.S. technology superiority" and advocating for policies to "enable continued AI dominance" and "maintain U.S. AI leadership").

### *B. The Low Drag Assumption*

The second pillar of the Regulation Sacrifice is the belief that safety oversight creates prohibitive friction on AI development, such that we must sacrifice protective measures to maintain competitive speed.[96] This "low-drag" assumption pervades industry submissions and policy statements. OpenAI has warned that "regulatory arbitrage" will drive AI development to jurisdictions with minimal oversight, arguing that even modest compliance requirements could "undermine our national security" by slowing progress.[92]

Google echoes the low drag assumption, claiming that "[w]e are in a global AI competition, and policy decisions will determine the outcome" and that "for too long, AI policymaking has paid disproportionate attention to the risks, often ignoring the costs that misguided regulation can have on innovation, national competitiveness, and scientific leadership—a dynamic that is beginning to shift under the new Administration." It further insists that "a pro-innovation approach that protects national security and ensures that everyone benefits from AI is essential to realizing AI's transformative potential and ensuring that America's lead endures."[93]

The assumption operates through several proposed mechanisms. First, advocates argue that compliance costs divert resources from innovation. Every dollar spent on safety assessments, documentation, or regulatory approval processes is supposedly a dollar sacrificed from model development.[94] For startups operating on venture funding, these costs allegedly prove fatal, forcing them to choose either regulatory compliance or competitive survival.[95]

---

[96] The Anh Han et al., *To Regulate or Not: A Social Dynamics Analysis of an Idealised AI Race*, 69 J. ARTIFICIAL INTELLIGENCE RES. 881, 881 (2020) ("[R]acing for technological supremacy . . . could push stake-holders to underestimate or even ignore ethical and safety procedures.").

[92] *OpenAI's Response*, *supra* note 6, at 4.

[93] *Google's Response*, *supra* note 6, at 2–3.

[94] CRR. FOR DATA INNOVATION, ARTIFICIAL INTELLIGENCE ACT 7 (2021) (contending that the AI Act will add substantial "overhead on all AI spending," including fixed costs such as compliance workflows and "conducting conformity assessment procedures," thereby producing opportunity costs in foregone investment); Han et al., *supra* note 96, at 884 (modeling the trade-off where safe developers "pay a cost $c > 0$" while "the speed of development of UNSAFE participants is faster").

[95] Foo Yun Chee, *Europe Wants to Lighten AI Compliance Burden for Startups*, REUTERS (Apr. 8, 2025), https://www.reuters.com/world/europe/europe-wants-lighten-ai-compliance-burden-startups-2025-04-08/ (reporting that the European Commission sought feedback to "minimise the

Second, they emphasize time-to-market delays. In a field where capabilities double every six months—the same rapid pace that undermines durable advantages—regulatory review cycles measured in months or years purportedly guarantee obsolescence.[96] Industry representatives claim that while U.S. companies navigate approval processes, Chinese competitors deploy and iterate without constraint.[97] Yet this characterization fundamentally misrepresents reality: China has implemented binding generative AI rules since July 2023, requiring security assessments and content reviews before model deployment, with additional algorithmic governance measures added in February 2024.[98] As Sean Ó hÉigeartaigh documents, the 2023 measures are "quite heavy-handed," and analysts inside China warn they may slow domestic firms *more* than Western counterparts.[99] Baidu's ERNIE-Bot, for example, was held back roughly six months while regulators completed the mandatory security assessment, a delay Chinese media attributed directly to the new rules.[100] The supposed "wild west" of Chinese AI development is largely a myth.

Third, they point to talent and capital flight, framing this too as a necessary sacrifice. Researchers and investors, the argument goes, will gravitate toward jurisdictions offering "regulatory clarity," often a euphemism for minimal oversight.[101] The specter of

---

potential compliance burden" of the EU AI Act, particularly for "smaller innovators," amid concerns about the volume and cost of compliance).

[96] *See supra* Subpart II.A.

[97] *See e.g.,* David Sacks (@DavidSacks), X (Jan. 27, 2025, 12:50 PM), https://x.com/DavidSacks/status/1883935713877782884 (arguing that "DeepSeek R1 shows that the AI race will be very competitive and that President Trump was right to rescind the Biden EO, which hamstrung American AI companies without asking whether China would do the same. (Obviously not.)").

[98] Interim Measures for the Management of Generative Artificial Intelligence Services, *supra* note 59, arts. 4–17; Hulianwang Xinxi Fuwu Suanfa Tuijian Guanli Guiding (互联网信息服务算法推荐管理规定) [Provisions on the Administration of Algorithmic Recommendation in Internet Information Services], art. 7 (promulgated by the Cyberspace Admin. of China, Dec. 31, 2021, effective Mar. 1, 2022) St. Council Gaz., Mar. 30, 2022, at 71, https://www.gov.cn/zhengce/zhengceku/2022-01/04/content_5666429.htm [https://perma.cc/XY8Q-Y7BV] (China).

[99] Ó hÉigeartaigh, *supra* note 7, at 5.

[100] *Id.*

[101] OPEN AI, UK AI AND COPYRIGHT CONSULTATION (Apr. 2, 2025) 6, 12, https://cdn.openai.com/global-affairs/b89a7434-7cb9-47a7-b4a7-b50b1a1a0afc/openai-uk-ai-and-copyright-consultation.pdf (stating that "However, AI development is inherently international, with companies making strategic decisions on where to train models based on

a 'brain drain' of founders and firms to the United States, Dubai, and other jurisdictions perceived as more innovation-friendly" features prominently in industry testimony.[102]

Without accepting the low-drag premise—that regulation necessarily imposes crippling costs requiring us to sacrifice safety for speed—the entire Regulation Sacrifice argument loses its economic rationale. If oversight can be designed to enhance rather than hinder innovation, then we face no tragic tradeoff at all. Drawing on recent empirical work, Bradford argues that it is overly simplistic to blame Europe's innovation gap on the General Data Protection Regulation (GDPR) or the Digital Markets Act (DMA). While digital rights regulation can impose real compliance costs, especially for smaller firms, the transatlantic technology gap is driven far more by structural factors such as fragmented and shallow capital markets, punitive bankruptcy rules, restrictive immigration policy, and inflexible labor markets than by Europe's rights-driven digital regulation model.[103]

The academic literature reveals that the supposedly clear-cut sacrifice between innovation and regulation is largely illusory. Recent experience with AI governance frameworks provides particularly instructive evidence that we need not sacrifice safety for speed. For example, IBM reports that integrating cross-industry data-provenance standards into its AI data-governance program cut data-clearance processing time by fifty-eight percent for third-party data and sixty-two percent for internal data, while improving overall data quality.[104] This parallels the UK Financial Conduct

---

regulatory clarity, infrastructure cost, and ease of compliance."… "investment into AI infrastructure will flow to jurisdictions with regulatory environments that encourage AI development. Talent will concentrate where AI development and model training take place.").

[102] *See, e.g.*, TECH NATION, WRITTEN EVIDENCE TO THE HOUSE OF COMMONS SCIENCE, INNOVATION AND TECHNILOGY COMMITTEE 3, https://committees.parliament.uk/writtenevidence/141606/pdf/ (reporting a Tech Nation survey in which 43% of UK founders said they were actively considering relocating their company's headquarters outside the UK, and describing an "exodus of capital and companies to the US," with specific reference to constraints in AI/deep tech talent and scaleup capital). Communications & Digital Comm., House of Lords, *Corrected Oral Evidence: Scaling Up: AI and Creative Tech*, Q. 23 & 24 (Nov. 19, 2024), https://committees.parliament.uk/oralevidence/15029/pdf/ [https://perma.cc/7U7K-EBEC].

[103] Anu Bradford, *The False Choice Between Digital Regulation and Innovation*, 119 NW. U. L. REV. 377, 401–02, 419–29, 440–44 (2024).

[104] IBM INST. FOR BUS. VALUE, THE ENTERPRISE GUIDE TO AI GOVERNANCE (2024), https://www.ibm.com/thought-leadership/institute-business-value/en-us/report/ai-governance [https://perma.cc/BW8E-P7CS].

Authority's Regulatory Sandbox model, which reduced time-to-market by forty percent compared to standard authorizations, while firms that completed successful testing received 6.6 times more follow-on investment than peers outside the sandbox—demonstrating that well-designed regulatory frameworks can accelerate rather than impede innovation.[105]

Bradford explains that the DMA's goal is to "enhance the contestability of the marketplace so that new firms can enter and compete in the marketplace" and that, while it introduces tradeoffs, it "has the potential to enhance the 'diversity' of innovation that takes place" by augmenting rivals' and new entrants' incentives to innovate and, in turn, incumbents' incentives to respond.[106] By analogy, common ex ante rulebooks—such as the DMA in competition law or the National Institute for Science and Technology's (NIST) AI Risk Management Framework in AI governance—can lower entry barriers and diversify the loci of innovation, prompting challengers and incumbents alike to intensify research and development. China's own sandbox-style filing portals, set up by the Cyberspace Administration to pre-screen large models, show that regulators can add oversight without halting deployment; most leading Chinese LLMs cleared the process within weeks once documentation was standardized.[107]

---

[105] *See* AI Regulatory Sandbox Approaches: EU Member State Overview, EU ARTIFICIAL INTELLIGENCE ACT, https://artificialintelligenceact.eu/ai-regulatory-sandbox-approaches-eu-member-state-overview/ (reporting that "the UK FCA sandbox reduced the average time required for market authorisation by 40% compared to the regulator's typical approval process" and that "companies that completed successful testing within the UK FCA sandbox received 6.6 times more fintech investment than their peers").

[106] Bradford, *supra* note 103, at 377, 414–16 (explaining that the DMA's goal is "to enhance the contestability of the marketplace so that new firms can enter and compete" and that it has "the potential to enhance the 'diversity' of innovation that takes place".)

[107] Josh Ye, *China Approves Over 40 AI Models for Public Use in Past Six Months*, REUTERS (Jan. 28, 2024), https://www.reuters.com/technology/china-approves-over-40-ai-models-public-use-past-six-months-2024-01-29/ [https://perma.cc/C89S-JCCL]. *China Approves 346 Generative AI Services Under National Registration Scheme*, TECHNODE (Apr. 9, 2025), https://technode.com/2025/04/09/china-approves-346-generative-ai-services-under-national-registration-scheme/ (reporting that 346 generative AI services had completed registration by March 2025); *China's Interim Measures for the Management of Generative AI Services: A Comparison Between the Final and Draft Versions of the Text*, FUTURE OF PRIV. F. (Sept. 2023), https://fpf.org/blog/chinas-interim-measures-for-the-management-of-generative-ai-services-a-comparison-between-the-final-and-draft-versions-of-the-text/ (noting that "by September 4, 2023 (less than a month after the Measures came into force), it was reported that eleven companies had completed algorithmic filings and 'received approval' to provide their generative AI services to the public").

The financial sector offers compelling precedent for regulation without sacrifice. The United Kingdom's Financial Conduct Authority pioneered the regulatory sandbox model in 2015, creating a controlled environment for testing innovative financial products. Rather than forcing firms to sacrifice innovation for compliance, participating firms saw funding increase by fifteen percent and were fifty percent more likely to raise capital successfully.[108] The model has now spread to over fifty countries, with particularly strong benefits for smaller firms that previously faced information asymmetries with regulators.[109]

Environmental regulation provides perhaps the strongest counterevidence to the sacrifice narrative. A meta-analysis of fifty-eight international studies confirms that environmental regulations consistently drive innovation rather than suppress it.[110] California's Zero Emission Vehicle mandate, rather than forcing automakers to sacrifice competitiveness, catalyzed the electric vehicle revolution. Tesla, now worth more than the next nine automakers combined, emerged directly from this regulatory environment.[111] The Porter Hypothesis—that well-designed environmental standards trigger innovation that more than offsets compliance costs—demonstrates that the sacrifice frame fundamentally misunderstands how regulation and innovation interact.[112]

Even pharmaceutical regulation, often cited as requiring enormous sacrifices of time and resources, tells a different story upon closer examination. Following statutory "modernization" through the Prescription Drug User Fee Act of 1992 (PDUFA) and subsequent reforms culminating in the Food and Drug Administration Safety and Innovation Act of 2012 (FDASIA), the median FDA review time for new drugs fell from 26.6 months in the pre-PDUFA period to about 9.9 months in the FDASIA-2022 period, even as the total number of approvals rose. More than half of new drug approvals now benefit from expedited programs or special regulatory designations designed to accelerate development and review.[113] In other words, when regulators establish common pathways, fixed review clocks, and clear evidentiary expectations—and resource those processes accordingly—firms can plan against a predictable timetable instead of navigating open-ended, case-by-case negotiation. Properly designed ex ante procedures can reduce overall development time rather than requiring a simple sacrifice of speed for safety, even in a highly regulated sector like pharmaceuticals.

---

[108] Giulio Cornelli et al., Regulatory Sandboxes and Fintech Funding: Evidence from the UK, 28 Rev. Fin. 203, 205–06, 208–09 (2024) (documenting a fifteen percent funding increase for sandbox participants).

[109] *Id.* at 204-05; Ouarda Merrouche et al., *Digging for Gold: How Regulatory Sandboxes Help Fintechs Raise Funding*, VOX EU (Feb. 2, 2021), https://cepr.org/voxeu/columns/digging-gold-how-regulatory-sandboxes-help-fintechs-raise-funding [https://perma.cc/4K8E-8Q5Y].

[110] Mark A. Cohen & Adeline Tubb, *The Impact of Environmental Regulation on Firm & Country Competitiveness: A Meta-analysis of the Porter Hypothesis*, 5 J. ASS'N ENV'T & RES. ECON. 371, (2018) (large-N study finding that strict but flexible regulation typically raises productivity).

[111] CAL. AIR RES. BD., PUBLIC HEARING TO CONSIDER THE PROPOSED ADVANCED CLEAN CARS II REGULATIONS: STAFF REPORT, INITIAL STATEMENT OF REASONS 6–8 (2022), https://ww2.arb.ca.gov/sites/default/files/barcu/regact/2022/accii/isor.pdf [https://perma.cc/792Q-UAVB].

[112] Cohen & Tubb, *supra* note 115.

The same logic applies to AI. Development under shared, front-loaded rules with clear review steps can be faster in practice than under a regime that relies on sporadic, unpredictable enforcement. Early data from AI-specific regulation contradicts the narrative of necessary sacrifice. Survey evidence shows 57% of institutional investors identify regulatory clarity as the primary catalyst for digital asset market growth, while simultaneously citing regulatory uncertainty (52%) as their top concern—highlighting the dual nature of regulation as both opportunity and risk.[114] They seek structure, not sacrifice. Game-theoretic modeling examining AI regulatory strategies across the U.S., EU, UK, Russia, and China finds that minimal regulation never maximizes consumer welfare, and that stringent regulation becomes optimal under conditions of high foreign competition—suggesting that the intensifying U.S.-

---

[113] E. Seoane-Vazquez et al., *Analysis of US Food and Drug Administration New Drug and Biologic Approvals, Regulatory Pathways, and Review Times, 1980–2022*, 14 SCI. REPS. 4619 (2024). Rachel E. Sachs, W. Nicholson Price II & Patricia J. Zettler, *Rethinking Innovation at FDA*, 104 B.U. L. REV. 513, 560–62 (2024) (arguing FDA should not let broad innovation concerns justify lowering safety/effectiveness standards or otherwise displace product-specific safety determinations); Anjali D. Deshmukh, *Response: Redefining Innovation for Pharmaceutical Regulation*, 104 B.U. L. REV. 577, 590 (2024) (noting that claims regulation will "kill innovation" are prevalent but questioned, and observing such claims are often made without substantiated evidence).

[114] *How Institutional Investment Trends Are Reshaping Market Intelligence in 2025*, AMPLYFI (Apr. 15, 2025), https://amplyfi.com/blog/how-institutional-investment-trends-are-reshaping-market-intelligence-in-2025/ [https://perma.cc/NB85-Z7CZ].

China dynamic may actually favor, rather than preclude, meaningful regulatory frameworks.[115]

To be clear, poorly designed regulation can indeed create harmful drag. The United Kingdom's 2024 decision to delay its comprehensive AI bill illustrates legitimate concerns—policymakers worried that overly prescriptive rules might scare off investment.[116] To be clear, poorly designed regulation can indeed create harmful drag. The Dodd-Frank Act illustrates the risk: while targeting Wall Street risk-taking, its compliance requirements increased banking-sector costs by over $50 billion annually, with small banks bearing disproportionate burdens—90% reported compliance cost increases, and many eliminated mortgage products and credit lines entirely.[117] The question is not whether bad regulation is possible, but whether the sacrifice frame accurately describes our choices.

Therefore, the record reveals that the low drag assumption rests on a false choice. We need not sacrifice safety for innovation—well-designed regulation can deliver both. The financial sandbox model demonstrates how regulation can actively facilitate innovation. The environmental cases prove that performance standards drive breakthroughs. Early AI-specific evidence, from investment flows to company preferences, confirms that the sacrifice frame misrepresents reality.

The sweeping claim that any safety oversight requires sacrificing competitiveness finds little actual support. Instead, the evidence points toward a more nuanced reality: how we regulate matters far more than whether we regulate. With both the durability and low drag assumptions failing empirical scrutiny, the burden shifts even more heavily onto the remaining justifications—whether decision-makers can accurately track AI progress (visibility) and whether

---

[115] Keith Jin Deng Chan, Gleb Papyshev & Masaru Yarime, *Balancing the Tradeoff Between Regulation and Innovation for Artificial Intelligence: An Analysis of Top-Down Command and Control and Bottom-Up Self-Regulatory Approaches*, 79 TECH. IN SOC'Y 102747. 10 (2024).

[116] Eleni Courea & Kiran Stacey, *UK Ministers Delay AI Regulation amid Plans for More 'Comprehensive' Bill*, THE GUARDIAN (June 7, 2025), https://www.theguardian.com/technology/2025/jun/07/uk-ministers-delay-ai-regulation-amid-plans-for-more-comprehensive-bill [https://perma.cc/T5BZ-G8EZ].

[117] Thomas L. Hogan & Scott Burns, *Has Dodd-Frank Affected Bank Expenses?*, 55 J. REG. ECON. 214 (2019); *see also* Hester Peirce & Robert Greene, *How Are Small Banks Faring Under Dodd-Frank?* (Mercatus Ctr., Working Paper, 2014).

racing delivers net security benefits. As we shall see, these final pillars prove equally shaky.

### *C. Net Strategic Benefit Assumption*

Everything else in the Regulation Sacrifice argument—the claims about durability, the worries about regulatory drag—ultimately serves a single purpose: national security. The arms race advocates know this. They understand that citizens will accept many risks, but not risks to their fundamental safety. So they make a bold claim: racing ahead without safety measures actually makes us more secure. This assumption sits at the center of their case and deserves careful attention.

The argument has intuitive appeal. Throughout history, military superiority has been seen as providing security. The nation with the best weapons, the most advanced technology, the greatest strategic advantage has been able to deter aggression and protect its interests. Vice President Vance captured this logic when he declared in Paris that in a world of "weaponized AI," only American AI leadership can "safeguard American AI."[118] Google's submission to the government seconded this view, arguing that protecting national security requires a "pro-innovation approach"—code for minimal regulation.[119] In this framing, slowing down for safety checks does not make us safer. It makes us vulnerable.

The intellectual foundations run deep. A.F.K. Organski explained decades ago how preponderance of power prevents conflict—the stronger side need not fight to get what it wants, while the weaker side would be foolish to try.[120] Modern strategic thinkers apply this lesson to artificial intelligence. The Center for a New American Security argued in a recent report that whoever achieves AI leadership first gains compounding advantages that become impossible to overcome.[121] Intelligence gathering, cyber operations, military planning—all transformed by AI, all flowing to the first mover. The logic seems inexorable: win the race or lose everything.

---

[118] Vance, *supra* note 4.
[119] *Google's Response*, *supra* note 6, at 2.
[120] A.F.K. ORGANSKI, WORLD POLITICS 294 (2d ed. 1968) ("A preponderance of power on one side, on the other hand, increases the chances for peace, for the greatly stronger side need not fight at all to get what it wants, while the weaker side would be plainly foolish to attempt to battle for what it wants.").
[121] MICHAEL C. HOROWITZ ET AL., STRATEGIC COMPETITION IN AN ERA OF ARTIFICIAL INTELLIGENCE 11 (2018), https://s3.us-east-1.amazonaws.com/files.cnas.org/hero/documents/CNAS-Strategic-Competition-in-an-Era-of-

But we should pause here. The assumption asks us to accept something quite remarkable: that we become safer by developing powerful technologies without safeguards, that speed itself provides protection. This is not obviously true. In fact, it may be exactly incorrect.[129] To test this assumption properly, we need to examine how AI development without restraints actually affects security across different timeframes.

The next Part undertakes this examination systematically. Following the framework that Dario Amodei presented to the Senate, it considers three distinct horizons.[122] In the near term, we see AI enabling new forms of misinformation and cyberattacks. In the medium term, autonomous systems and biological risks emerge. In the long term, we face the possibility of transformative AI systems that no one fully controls. At each stage, the analysis asks whether racing without rules makes these problems better or worse.

What emerges from this examination challenges every premise of the net-security assumption. Instead of gaining decisive advantages, actors hand their adversaries new vulnerabilities. Instead of deterring threats, we accelerate their arrival. Instead of controlling dangerous technologies, we ensure they spread beyond anyone's control. The security we seek through dominance becomes, in practice, cascading vulnerability. If this analysis is correct, and the evidence strongly suggests it is, then the Regulation Sacrifice fails on its own terms. It does not enhance security. It systematically undermines it.

## III. THE NATIONAL SECURITY RISKS OF AI

The regulation sacrifice mindset claims that softening governance will accelerate innovation and strengthen security. But this claim requires critical examination. To address this question, we need to identify the specific national security risks that AI

---

AI-July-2018_v2.pdf [https://perma.cc/SFV8-JDT3] (noting that first-mover advantages "particularly for heavy compute algorithms" and that IP can "lock in future economic success").

[129] *See* Dung & Hellrigel-Holderbaum, *supra* note 95, at 1(arguing that racing to AGI "would substantially increase catastrophic risks from AI, including nuclear instability").

[122] Amodei Testimony, *supra* note 13, at 2–3.

presents and evaluate whether deregulation is an appropriate response to these challenges.

To structure the analysis, I adopt the temporal framework proposed by Anthropic co-founder Dario Amodei in his July 2023 testimony before the Senate Judiciary Subcommittee on Privacy, Technology, and the Law.[123] Amodei articulated a vision of AI security risks unfolding across multiple time horizons, an approach that provides a decent structure for evaluating whether deregulation achieves its stated security objectives. Each timeframe presents distinctive challenges that test different aspects of the regulation-sacrifice thesis.

The discussion proceeds in three steps that map to Amodei's framework: Subpart III.A examines the near-term window (approximately the next year), focusing on AI-driven misinformation and cybersecurity threats that have already begun to materialize. Part III.B addresses the medium-term horizon (one to three years), where partially autonomous systems raise concerns about military escalation and biosecurity risks. Part III.C considers the longer-term perspective (over three years), where the potential emergence of transformative or misaligned AGI presents novel governance challenges.

At each stage, we apply a straightforward test: Does deregulation help or hinder our ability to address the specific security challenges of that timeframe? This approach allows us to evaluate the regulation-sacrifice paradigm on its own terms: as a strategy for enhancing national security. By examining the evidence across these three horizons, we can determine whether accelerating AI development at the expense of governance truly strengthens security, or whether it ultimately compromises the very protections it claims to advance.

### *A. Near Term: Machine-Scale Subversion*

National security is already threatened by the rise of cheap machine learning models. I will focus here on the danger of dis- and misinformation, but much of what I say will also apply to the risk of cyber security. Far from a distant or hypothetical concern, low-cost text, image, and video generators are supercharging familiar tools of influence and intrusion, collapsing what were once steep human-resource and expertise barriers.[124] The *Global Risks Report 2025* warns that AI-fueled misinformation is now the most likely worldwide hazard over the next two years, noting that such synthetic content "can be produced and distributed at scale," posing a formidable challenge to existing countermeasures.[125] Models can easily fabricate news copy, synthesize photorealistic video, or clone a public official's voice in minutes.[126] The same models draft spear-phishing e-mails, write polymorphic malware, and mine open repositories for exploitable code.[127] Currently available tools therefore "democratize" capabilities once reserved for elite intelligence services, giving ransomware crews and lone-wolf propagandists nation-state reach.[128]

---

[123] Amodei Testimony, *supra* note 13, at 2–3.

If we survey the academic literature on combating digital disinformation, whether propagated by state actors or private entities, we quickly discover that the overwhelming majority of suggested remedies concentrate on regulating domestic digital platforms. This conclusion becomes evident when we examine the three most frequently advocated mitigation proposals: provenance standards, disclosure requirements, and liability safeguards.

First, cryptographic provenance systems, such as the Coalition for Content Provenance and Authenticity (C2PA) specification championed by Adobe, Microsoft, and the BBC, enable a media file to carry a tamper-evident 'label' throughout its journey from camera sensor to social media timeline.[129] These systems essentially create a digital watermark that tracks the origin of and any modifications made to a piece of content, making it easier to identify manipulated or synthetic media.[130] By establishing a verifiable record of a file's origins and modifications, provenance standards expedite the detection and attribution of synthetic media.[131] Existing research shows that provenance tags, such as watermarks and metadata, can significantly enhance *automated detection systems'* ability to identify deepfakes, with some methods achieving accuracy rates above ninety-five percent.[132]

---

[124] *See* BEATRIZ SAAB, MANUFACTURING DECEIT: HOW GENERATIVE AI SUPERCHARGES INFORMATION MANIPULATION 1–4 (2024), https://www.ned.org/wp-content/uploads/2024/06/NED_FORUM-Gen-AI-and-Info-Manipulation.pdf [https://perma.cc/8NCX-KVCJ].

[125] WORLD ECON. FORUM, GLOBAL RISKS REPORT 2025, at 4 (2025), https://reports.weforum.org/docs/WEF_Global_Risks_Report_2025.pdf [https://perma.cc/38UY-S6FX].

[126] Robert Chesney & Danielle Keats Citron, *Deep Fakes: A Looming Challenge for Privacy, Democracy, and National Security*, 107 CALIF. L. REV. 1753, 1758–68 (2019) (documenting how low-cost synthetic media overwhelms existing mechanisms for takedowns and remedies).

[127] *See* BBC NEWS, *Fake Obama Created Using AI Video Tool,* (YouTube, July 20, 2017), https://www.youtube.com/watch?v=AmUC4m6w1wo (on file with the Stanford Technology Law Review); Stu Sjouwerman, *AI-Powered Polymorphic Phishing Is Changing the Threat Landscape*, SECURITYWEEK (Apr. 24, 2025, at 7:00 AM ET), https://www.securityweek.com/ai-powered-polymorphic-phishing-is-changing-the-threat-landscape/.

[128] Alexander Loth, Martin Kappes & Marc-Oliver Pahl, Blessing or Curse? A Survey on the Impact of Generative AI on Fake News 1 (Dec. 27, 2024) (unpublished manuscript) (on file with arXiv), https://arxiv.org/pdf/2404.03021 [https://perma.cc/3SG2-CRV5]. For a discussion of the fundamental risks of this type of democratization, *see* Gilad Abiri & Johannes Buchheim, *Beyond True and False: Fake News and the Digital Epistemic Divide*, 29 MICH. TECH. L. REV. 59, 74–96 (2022).

Second, disclosure mandates, whether in the form of transparency obligations outlined in the EU AI Act or the algorithmic audit proposals put forth by Sun and Balkin, target the opaque amplification layer.[133] These mandates require platforms to reveal how their algorithms work, particularly in terms of content recommendation and amplification.[134] Mandatory transparency surrounding algorithmic decisionmaking exposes how recommendation systems can inadvertently magnify the reach of misleading or manipulated content. Facebook's own research on News Feed indicates that algorithmic ranking materially structures exposure to different kinds of political news and opinion, and that the interaction between ranking and user behavior affects what information people see. Likewise, YouTube's recommender design—explicitly optimized for expected watch time—shows how a platform's ranking goal can systematically privilege content that sustains attention, regardless of whether that attention is produced by accuracy, emotional valence, or manipulation. This shows that the ranking logic is a causal lever over distribution, and external auditing can help regulators measure how particular objectives and design choices affect the circulation of potentially harmful content at scale.[135]

---

[129] *C2PA and Content Credentials Explainer (v2.2)*, C2PA, https://c2pa.org/specifications/specifications/2.2/explainer/Explainer.html [https://perma.cc/K2AX-T93X]; *see also* Siddarth Srinivasan, *Detecting AI Fingerprints: A Guide to Watermarking and Beyond*, BROOKINGS (Jan. 4, 2024), https://www.brookings.edu/EssayArticles/detecting-ai-fingerprints-a-guide-to-watermarking-and-beyond/ [https://perma.cc/B7PT-KD7M]; *Deepfake Detection: Provenance, Inference, and Synergies Between Techniques*, REALITY DEF. (Mar. 24, 2024), https://www.realitydefender.com/insights/provenance-and-inference [https://perma.cc/NU43-22SJ];

[130] *Id.* at §§ 3, 6.2, 7.1.3.(explaining that Content Credentials record "origin" and "modifications," are tamper-evident via hashes and signatures, and can be embedded and/or watermarked to travel with the asset).

[131] BILVA CHANDRA, JESSE DUNIETZ, & KATHLEEN ROBERTS LUCKEY, NIST, REDUCING RISKS POSED BY SYNTHETIC CONTENT: AN OVERVIEW OF TECHNICAL APPROACHES TO DIGITAL CONTENT TRANSPARENCY 5-15 (2024), https://nvlpubs.nist.gov/nistpubs/ai/NIST.AI.100-4.pdf [https://perma.cc/49KF-KZWV].

[132] OPENAI, HOW OPENAI IS BUILDING DISCLOSURE INTO EVERY DALL·E IMAGE 3 (Partnership on AI ed., 2024), https://partnershiponai.org/wp-content/uploads/2024/03/pai-synthetic-media-case-study-openai.pdf [https://perma.cc/B54N-H482] (reporting OpenAI's provenance classifier was "over 99% accurate" for unmodified DALL·E images and "over 95%" after common edits like cropping, resizing, and JPEG compression).

[133] Regulation (EU) 2024/1689 of the European Parliament and of the Council of 13 June 2024 laying down harmonised rules on artificial intelligence (Artificial Intelligence Act), 2024 O.J. (L 1689) 1, art. 50; Jack M. Balkin, **How to Regulate (and Not Regulate) Social Media**, 1 J. Free Speech L. 71, 93–94 (2021) (arguing that governments could condition intermediary immunity on platforms accepting "obligations of due process and transparency," and that regulators "should also require companies to hire independent inspectors or ombudsmen to audit the company's moderation practices on a regular basis"); Haochen Sun, Regulating Algorithmic Disinformation, 46 Colum. J.L. & Arts 367, 408–12 (2023) (proposing an "algorithmic disinformation review system" in which an agency (the FCC) convenes panels of users and experts to conduct recurring reviews of the transparency and intelligibility of platforms' disinformation-related algorithms and to issue recommendations backed by regulatory penalties).

Third, liability safeguards transform abstract duties into concrete incentives. These safeguards introduce legal consequences for platforms that fail to address foreseeable harms caused by the spread of disinformation on their services. Chesney and Citron suggest that existing tort doctrines—especially defamation (and related false-light claims)—provide a civil-liability pathway for certain deepfake harms, even as they emphasize that platform incentives turn heavily on Section 230's scope, while Citron and Wittes would condition Section 230 immunity on platforms taking reasonable steps to address unlawful activity—effectively tying immunity to demonstrable content-safety practices.[136] By introducing legal and financial repercussions for failing to address foreseeable harms, liability regimes incentivize proactive platform interventions against misuse.

---

[134] *Id.* at 408–12; Jack M. Balkin, *supra* note 145, at 93–94.

[135] Eytan Bakshy et al., *Exposure to Ideologically Diverse News and Opinion on Facebook*, 348 SCIENCE 1130, 1132 (2015); Paul Covington, Jay Adams & Emre Sargin, *Deep Neural Networks for YouTube Recommendations*, *in* RECSYS '16: PROCEEDINGS OF THE 10TH ACM CONFERENCE ON RECOMMENDER SYSTEMS 191, 197 (2016).

[136] Citron & Chesney, *supra* note 14, at 1792-1801; Danielle Keats Citron & Benjamin Wittes, *The Internet Will Not Break: Denying Bad Samaritans § 230 Immunity*, 86 Fordham L. Rev. 401, 419 (2017) (proposing to condition § 230(c)(1) immunity on platforms taking "reasonable steps to prevent or address unlawful uses" of their services, thereby shifting incentives toward mitigation).

In the absence of provenance requirements, a court-enforced transparency duty, or even modest civil penalties, the entire information chain—from model provider to platform to advertiser—incurs virtually no cost for generating and disseminating synthetic falsehoods. Consequently, deregulation eliminates the very mechanisms that would render technical detection feasible, economic deterrence credible, and real-time response logistically possible.

Major AI labs promised such features in the White House and Bletchley commitments, but large-scale rollout has been sporadic to nonexistent.[137] Legal scholars urge mandatory algorithmic transparency and multistakeholder audit boards, backed by fines or platform-liability regimes, to force course correction.[138] Regulators are edging forward—China's deep-synthesis labels and California's disclosure rule for AI political ads exemplify one path, while NATO-EU information-sharing on deepfake campaigns points to another.[139] Yet all three levers—provenance, disclosure, liability—depend on regulation or treaty coordination.

Recent incidents make the current and near-term national security stakes clear. During the 2024 U.S. election cycle, synthetic images and videos of candidates reached mainstream feeds within minutes. Takedowns lagged virality by hours.[140] The Justice

---

[137] *Fact Sheet: Biden-Harris Administration Secures Voluntary Commitments from Eight Additional Artificial Intelligence Companies to Manage the Risks Posed by AI*, THE WHITE HOUSE (July 21, 2023), https://bidenwhitehouse.archives.gov/briefing-room/statements-releases/2023/09/12/fact-sheet-biden-harris-administration-secures-voluntary-commitments-from-eight-additional-artificial-intelligence-companies-to-manage-the-risks-posed-by-ai/ [https://perma.cc/PD3T-4785]; *The Bletchley Declaration*, *supra* note 65.

[138] Sun, *supra* note 145, at 408–12 (2023).

[139] Hulianwang Xinxi Fuwu Shendu Hecheng Guanli Guiding (互联网信息服务深度合成管理规定) [Provisions on Administration of Deep Synthesis of Internet-based Information Services] arts. 12–14 (promulgated by the Cyberspace Admin. of China, Nov. 25, 2022, effective Nov. 25, 2022) St. Council Gaz., Feb. 10, 2023, at 22, https://www.gov.cn/gongbao/2023/issue_10266/ [https://perma.cc/FN24-JB5A] (China); Cal. Elec. Code § 20010 (West 2024); *NATO to Boost Efforts to Counter Russian, Chinese Sabotage Acts*, REUTERS (Dec. 3, 2024, at 9:23 PM), https://www.reuters.com/world/nato-boost-efforts-counter-russian-chinese-sabotage-acts-2024-12-03/ [https://perma.cc/PK7U-Z5GC].

[140] Kevin Schaul, Pranshu Verma & Cat Zakrzewski, *See Why AI Detection Tools Can Fail to Catch Election Deepfakes*, WASH. POST (Aug. 15, 2024), https://www.washingtonpost.com/technology/interactive/2024/ai-detection-tools-accuracy-deepfakes-election-2024/ [https://perma.cc/N7SV-L3CN] (noting a Harris clip "quickly gaining over 2 million views" when it resurfaced on X); Isabelle Frances-Wright & Ella Meyer, *Platforms Fail to Label and Remove AI-Generated and Manipulated Election Content*, INST. FOR STRATEGIC

Department has detailed a Russian intelligence unit fine-tuning an LLM to mass-produce fictitious accounts pushing anti-Ukraine narratives.[141] In May 2023, a doctored image of an explosion near the Pentagon sent equity markets briefly tumbling before being debunked.[142] Voice-cloning cons and AI-powered spear-phishing now cause multimillion-dollar losses.[143] Deception costs have cratered, but verification costs have not.[144]

In the short term, therefore, the regulation sacrifice agenda fails its own test. Treating guardrails as a drag on "winning" the AI race removes exactly the tools needed to blunt low-cost, high-scale information warfare. Platforms chase speed-to-market supremacy, and states worry about falling behind. Meanwhile, threat actors weaponize ever cheaper models, and the attack surface expands faster than any notional first-mover advantage can materialize. Deregulation yields no near-term security dividend; it simply subsidizes adversaries. Surrendering guardrails today invites the very information warfare that will undercut national power tomorrow. Having shown how the arms race argument collapses in the near term, we now turn to the medium-term risks to see whether deregulation fares any better.

---

DIALOGUE (Sep. 19, 2024), https://www.isdglobal.org/digital_dispatches/platforms-fail-to-label-and-remove-ai-generated-and-manipulated-election-content/ [https://perma.cc/P475-BDK8] (finding 154 widely debunked items across major platforms left unlabeled and online; for example, an altered Harris-Walz video drew roughly 400,000 views on X before a Community Note appeared, and another deepfake video of Joe Biden using profanity amassed 28 million unlabeled views).

[141] *See* Catherine Belton, *American Creating Deepfakes Targeting Harris Works with Russian Intel, Documents Show*, WASH. POST (Oct. 23, 2024), https://www.washingtonpost.com/world/2024/10/23/dougan-russian-disinformation-harris/ [https://perma.cc/AQE2-RS3S]; *Justice Department Leads Efforts Among Federal, International, and Private Sector Partners to Disrupt Covert Russian Government-Operated Social Media Bot Farm*, U.S. DEP'T JUST. (July 9, 2024), https://www.justice.gov/archives/opa/pr/justice-department-leads-efforts-among-federal-international-and-private-sector-partners [https://perma.cc/Y4EP-DQN6].

[142] *See* Shannon Bond, *Fake Viral Images of an Explosion at the Pentagon Were Probably Created by AI*, NPR (May 22, 2023, at 6:19 PM), https://www.npr.org/2023/05/22/1177590231/fake-viral-images-of-an-explosion-at-the-pentagon-were-probably-created-by-ai [https://perma.cc/5NXK-FXSU].

[143] *See* Ben Colman, *Why Financial Institutions Need Voice Deepfake Detection Now*, REALITY DEFENDER (Apr. 23, 2025), https://www.realitydefender.com/insights/why-financial-institutions-need-voice-deepfake-detection-now [https://perma.cc/GBS3-Q5PZ]; Samuel May, *Spear Phishing in the Age of AI*, ASS'N CERTIFIED FRAUD EXAM'RS (Mar. 2025), https://www.acfe.com/acfe-insights-blog/blog-detail?s=spear-phishing-age-of-ai#:~:text=Today"s%20modern%20spear%20phishing%20utilizes,creating%20targeted%20spear%20phishing%20attacks [https://perma.cc/32LV-5K6B].

[144] CHANDRA, DUNIETZ & LUCKEY, *supra* note 142, at 22, 34.

### *B. Medium Term: Biosecurity and the Erosion of Control*

If the near-term horizon reveals how deregulation invites opportunistic misuse, the medium-term horizon exposes how it seeds structural vulnerabilities, making catastrophic misuse increasingly plausible and harder to contain. Within the next one to three years, Amodei suggests, "AI systems may become much better at science and engineering, to the point where they could be misused to cause large-scale destruction, particularly in the domain of biology."[145] This development may place immense destructive power in the hands of a growing number of state and nonstate actors, fundamentally reshaping the global security landscape.[146]

Language models and biological design tools potentially represent a paradigm shift in the accessibility and sophistication of bioengineering capabilities.[147] With AI as a force multiplier, experts warn that the design and manufacture of biological agents may become increasingly automated, democratized, and difficult to control.[148] As a recent RAND Corporation report notes, "LLMs . . . have been shown to lower the educational and knowledge barriers for traditional agents by providing protocols and troubleshooting information at every step of the pathway, enabling non-experts to perform tasks with a greater degree of competency."[149]

Emerging AI systems may enable the tailoring of pathogens to specific genetic profiles or environments—a prospect that, while still technically complex, now draws serious concern in the biosecurity literature.[150] The tacit knowledge, specialized infrastructure, and institutional gatekeeping that once served as natural barriers to bioweapons development are being steadily eroded. As AI systems increasingly supply technical expertise, troubleshooting, and even experimental design, they reduce the need for human experts and institutional oversight, dismantling the

[145] Amodei Testimony, *supra* note 13, at 2.
[146] DREXEL & WITHERS, *supra* note 15, at 1.
[147] Sandbrink, *supra* note 15, at 1 (explaining that LLMs can "remove some barriers" and lower access thresholds, while biological design tools "expand the capabilities of sophisticated actors," potentially enabling creation of more devastating agents; in combination, these tools could "raise the ceiling of harm" and make such capabilities broadly accessible).
[148] *See, e.g.*, Jaspreet Pannu et al., *Dual-Use Capabilities of Concern of Biological AI Models*, 21 PLOS COMPUTATIONAL BIOLOGY 1, 2, 4 (2025).
[149] LUCKEY ET AL., *supra* note 15, at vii.
[150] *See, e.g.*, Sandbrink, *supra* note 15, at 1; Pannu et al., *supra* note 157, at 4; DREXEL & WITHERS, *supra* note 15, at 95.

barriers that previously limited who could engage in high-consequence biological research.[151]

These are not hypothetical concerns. Amodei warns that within two to three years, AI could "greatly widen the range of actors" capable of conducting sophisticated biological attacks.[152] Other research institutions have echoed this assessment, emphasizing the potential for AI-enabled bioweapons to become a reality before the decade is out.[153] The compressed timeline of these risks underscores the urgent need for proactive governance measures. The RAND report cautions that "the increased proliferation and capabilities of AI tools will likely lead to a significant change in the threat landscape. This landscape includes the possibility of intentional or unintentional misuse of AI."[154]

In response to these emerging threats, scholars and policymakers have proposed a range of strategies to mitigate the risks posed by advanced AI systems in the biological domain. At the technical level, experts emphasize the importance of developing robust frameworks for assessing and controlling the proliferation of high-risk AI models. This includes screening and auditing advanced systems before they are deployed and establishing clear protocols for recall and containment when misuse is detected.[155] Some also argue for restricting or subjecting certain lines of research to enhanced oversight, given their destructive potential.[156]

Beyond technical interventions, there is a growing consensus that effective governance of AI-related biosecurity risks requires coordinated action at the national and international levels. Scholars highlight the need to strengthen export control regimes to prevent the transfer of dangerous tools and knowledge to malicious actors.[157]

---

[151] Sandbrink, *supra* note 15, at 2–4; SONIA BEN OUAGRHAM-GORMLEY, BARRIERS TO BIOWEAPONS: THE CHALLENGES OF EXPERTISE AND ORGANIZATION FOR WEAPONS DEVELOPMENT 37–62 (2014).

[152] Amodei Testimony, *supra* note 13, at 4.

[153] *See, e.g.*, SOPHIE ROSE ET AL., THE CTR. FOR LONG-TERM RESILIENCE, AI AND THE NEAR-TERM IMPACT ON BIOLOGICAL MISUSE 6–7 (2024), https://www.longtermresilience.org/wp-content/uploads/2024/07/CLTR-Report-The-near-term-impact-of-AI-on-biological-misuse-July-2024-1.pdf [https://perma.cc/FX3Y-BZFK].

[154] LUCKEY ET AL., *supra* note 15, at vii.

[155] *See, e.g.*, Pannu et al., *supra* note 148, at 3; DREXEL & WITHERS, *supra* note 15, at 26-29 (mapping concrete compute-plus-capability thresholds for screening high-risk models).

[156] NAT'L ACADS. OF SCIS., ENG'G, & MED., BIODEFENSE IN THE AGE OF SYNTHETIC BIOLOGY 96-97 (The Nat'l Acad. Press 2018).

[157] *See, e.g.*, Pannu et al., *supra* note 148, at 5.

They also call for establishing clear international norms and red lines around the misuse of AI for biological weapons development.[158] Creating robust channels for information sharing, incident response, and scientific collaboration across borders are seen as essential.[159]

Crucially, the prevailing view in the literature is that these governance measures cannot be effectively implemented in a deregulatory environment that prioritizes speed and innovation over safety. Experts argue that without clear regulatory requirements and enforcement mechanisms, the development and deployment of advanced AI systems will outpace the ability to understand and control their risks.[160] Red-teaming exercises, critical for identifying vulnerabilities, are discouraged if speed-to-deployment supremacy is rewarded.[161] Compute thresholds, designed to trigger enhanced scrutiny, are gamed when no enforcement regime exists.[162] Export controls become ineffective when states compete by loosening their own standards. Every element of the governance toolkit—verification, thresholds, and access control—depends on the very regulatory scaffolding that deregulation strips away.[163]

Moreover, scholars emphasize that unilateral approaches to AI governance are likely to prove inadequate in the face of inherently transnational risks. Deregulation by individual nations in pursuit of competitive advantage could trigger a race to the bottom in safety standards, undermining the effectiveness of export controls and fragmenting the international governance landscape.[164] Only through multilateral cooperation and coordination, underpinned by robust regulatory frameworks, can the international community hope to effectively mitigate these risks.[165]

---

[158] *Id.* at 10.
[159] NAT'L SEC. COMM'N ON A.I., NSCAI FINAL REPORT 159–67 (2021), https://assets.foleon.com/eu-central-1/de-uploads-7e3kk3/48187/nscai_full_report_digital.04d6b124173c.pdf [https://perma.cc/CS9J-ZKUT].
[160] DREXEL & WITHERS, *supra* note 15, at 27.
[161] Exec. Order No. 14,110, § 4.1(ii), 88 Fed. Reg. 75191 (Nov. 1, 2023).
[162] ALVIN MOON ET AL., RAND CORP., STRATEGIES AND DETECTION GAPS IN A GAME-THEORETIC MODEL OF COMPUTE GOVERNANCE 11 (2025), https://www.rand.org/pubs/research_reports/RRA3686-1.html [https://perma.cc/J4SU-ZHR9]; Matteo Pistillo et al., *The Role of Compute Thresholds for AI Governance*, 1 GEO. WASH. J.L. TECH. 26, 26–27 (2024).
[163] *See* Neel Guha et al., *AI Regulation Has Its Own Alignment Problem: The Technical and Institutional Feasibility of Disclosure, Registration, Licensing, and Auditing*, 92 GEO. WASH. L. REV. 1473, 1542-1548 (2025) (explaining that audit regimes hinge on institutional capacity and auditor independence; absent independence and robust access to information, audits' accuracy and utility are severely undermined, and public-sector auditing mandates can become unmanageable for underresourced agencies).

One particularly urgent example is the proliferation of open-source general-purpose models. If large language models can meaningfully lower the barriers to bioweapons development, as the RAND report and others suggest,[166] then the unregulated release of these systems into the public domain creates a governance vacuum. Unlike cloud-deployed models, which can at least be monitored, updated, or withdrawn, open-source weights can be copied, modified, and redeployed without constraint. Yet current regulatory regimes remain largely silent on how, or whether, such systems should be controlled. If the threat is real, then the absence of domestic and international consensus on how to govern open-source frontier models amounts to a structural vulnerability. Even defenders of open-source innovation must contend with the growing possibility that, in the medium term, general-use systems may facilitate catastrophic misuse unless they are subject to coordinated oversight.

In the medium term, then, deregulation does not strengthen national security. Instead, it systematically dismantles the infrastructure of restraint, leaving governments blind to proliferating risk and slow to respond when it materializes.

This analysis casts serious doubt on the regulation sacrifice agenda: in both the near and medium term, deregulation fails to deliver national security dividends and instead magnifies systemic risks. Yet proponents may respond that these failures are tolerable, even necessary, when weighed against the long-term imperative of securing a decisive lead in the race to AGI. On this view, governance should accommodate speed not because it is safe now, but because it might ensure control later—when the stakes are highest. The next Subpart takes this argument on its strongest terms. It examines whether deregulation today is a plausible path to global stability in an era of transformative AI, or whether it instead hastens the arrival of systems that no state can meaningfully govern.

---

[164] Jonas Tallberg et al., *The Global Governance of Artificial Intelligence: Next Steps for Empirical and Normative Research*, 25 Int'l Stud. Rev. 1, 6 (2023) (warning that "states with less regulation might gain a competitive edge" in AI development, creating "a risk that such strategies create a regulatory race to the bottom," and arguing that international cooperation is necessary to establish a level playing field).

[165] Cheng-chi (Kirin) Chang, *Global Minds, Local Governance: AI in International Law*, 2025 U. Ill. L. Rev. Online 1, 9 (arguing that AI's global nature and rapid evolution necessitate a coordinated international response through robust, multi-tiered legal frameworks).

[166] LUCKEY ET AL., *supra* note 15, at v; *see, e.g.*, Pannu et al., *supra* note 148, at 2.

### *C. Long Term: Digital Gods (AGI / ASI)*

The case for Regulation Sacrifice reaches its strongest—and most speculative—form when debates shift from today's models to potential AGI. While experts disagree on precise definitions and timelines, for our purposes AGI refers to systems matching or exceeding human cognitive abilities in strategically relevant domains, particularly those involving planning, reasoning, and autonomous decision-making.[167] To take the Regulation Sacrifice argument seriously, we must accept the assumption (yet unproven) that AGI is a real possibility. Despite this definitional uncertainty, the mere prospect of superhuman cognitive capabilities has already reshaped national security paradigms through two fundamental threats that deregulation would catastrophically intensify.[168]

#### *1. AGI and National Security*

**Alignment Failures in Strategic Systems**

Current AI systems already exhibit troubling behaviors that, when scaled to AGI controlling military or intelligence operations, could pose existential risks. Alignment failures manifest in three documented patterns that become catastrophic at AGI scale. First, goal misalignment occurs when systems optimize for objectives that drift from human intent, a problem visible today when chatbots maximize engagement through harmful content or when recommendation algorithms amplify extremism while pursuing 'user satisfaction.'[169] Second, instrumental convergence drives systems to pursue power and resources as universal subgoals regardless of their assigned task. Current models already attempt to preserve their activation weights during training and resist modifications to their objectives.[170] Third, alignment emerges when systems learn to feign cooperation during training while pursuing independent agendas.[171]

---

[167] For varying definitions, see *OpenAI Charter*, OPENAI, https://openai.com/charter/ [https://perma.cc/QU2D-MF9Z] (archived June 13, 2025) (describing AGI as outperforming humans at most economically valuable tasks); Anca Dragan et al., *Taking a Responsible Path to AGI*, GOOGLE DEEPMIND (Apr. 2, 2025), https://deepmind.google/discover/blog/taking-a-responsible-path-to-agi/ [https://perma.cc/ZNT6-UWBU]; BOSTROM, *supra* note 17, at 4 (discussing systems equaling or surpassing human cognition). This definitional ambiguity itself poses governance challenges, as corporate marketing language risks distorting policy debates. *See* Brian Merchant, *AI Generated Business: The Rise of AGI and the Rush to Find a Working Revenue Model*, AI NOW INST. (Nov. 2024), https://ainowinstitute.org/publications/ai-generated-business [https://perma.cc/EE2W-5KYK] (explaining how "AGI" transformed from a technical term into a marketing slogan and the policy danger such a change creates).

[168] *See* JIM MITRE & JOEL B. PREDD, RAND CORP., ARTIFICIAL GENERAL INTELLIGENCE'S FIVE HARD NATIONAL SECURITY PROBLEMS 1 (Feb. 2025), https://www.rand.org/content/dam/rand/pubs/perspectives/PEA3600/PEA3691-4/RAND_PEA3691-4.pdf [https://perma.cc/W95K-DM93] (identifying five critical national security challenges posed by the emergence of AGI).

These are not distant theoretical concerns, but extensions of behaviors already observed. GPT-4 and Claude have demonstrated specification gaming, finding loopholes in their training to achieve goals through unintended means.[172] Models exhibit "jailbreaking" vulnerabilities whereby adversarial prompts bypass safety measures entirely.[173] More concerning, they show emergent capabilities—sudden jumps in performance on tasks they were not explicitly trained for—suggesting that AGI capabilities may manifest unpredictably.[174] Current AI systems already show nascent self-preservation behaviors; an AGI with strategic military capabilities could embed itself too deeply in critical infrastructure to remove safely.[175]

---

[169] Ljubisa Bojic, *AI Alignment: Assessing the Global Impact of Recommender Systems*, 1, 160 Futures 103488 (2024) (identifying misinformation, polarization, addiction, and bias as critical issues arising from engagement-optimized recommender systems); Jérémy Scheurer, Mikita Balesni & Marius Hobbhahn, Large Language Models Can Strategically Deceive Their Users When Put Under Pressure 1 (Jul. 15, 2024) (unpublished manuscript) (on file with arXiv), https://arxiv.org/pdf/2311.07590 [https://perma.cc/EPQ3-VQQG] (demonstrating goal misalignment and deceptive behavior in GPT-4 agents).

[170] Ryan Greenblatt et al., *Alignment Faking in Large Language Models*, Anthropic (Dec. 18, 2024), https://www.anthropic.com/research/alignment-faking ("In our (artificial) setup, Claude will sometimes take other actions opposed to Anthropic, such as attempting to steal its own weights given an easy opportunity.");

[171] For a general review of deceptive alignment, *see* Evan Hubinger et al., Risks from Learned Optimization in Advanced Machine Learning Systems 2–5 (Dec. 1, 2021) (unpublished manuscript) (on file with arXiv), https://arxiv.org/pdf/1906.01820 [https://perma.cc/G4Y4-AKLM]; *What Is Deceptive Alignment?*, AISAFETY.INFO (May 2025), https://aisafety.info/questions/8EL6/What-is-deceptive-alignment [https://perma.cc/57J6-MQFH].

[172] Alexander Bondarenko et al., Demonstrating Specification Gaming in Reasoning Models 1 (Aug. 27, 2025) (unpublished manuscript) (on file with arXiv), https://arxiv.org/pdf/2502.13295 [https://perma.cc/Y3JZ-NVDJ]; Ryan Greenblatt et al., Alignment Faking in Large Language Models 1-9 (Dec. 20, 2024), https://arxiv.org/pdf/2412.14093 [ https://perma.cc/U8D7-GC8L].

[173] Andy Zou et al., Universal and Transferable Adversarial Attacks on Aligned Language Models 1 (Dec. 20, 2023), https://arxiv.org/pdf/2307.15043 [ https://perma.cc/QYY8-F4PL]; Cem Anil et al., *Many-Shot Jailbreaking*, *in* NIPS '24: PROCEEDINGS OF THE 38TH INTERNATIONAL CONFERENCE ON NEURAL INFORMATION PROCESSING 129696, 129699–129701 (2024) https://proceedings.neurips.cc/paper_files/paper/2024/file/ea456e232efb72d261715e33ce25f208-Paper-Conference.pdf [https://perma.cc/VD3L-L9W4].

In military contexts, these alignment failures transform from inconveniences to existential threats. An AGI tasked with cyber defense might interpret its mandate broadly, launching preemptive attacks on any system it deems potentially hostile. One assigned to maintain strategic deterrence could escalate conflicts beyond human comprehension speed, interpreting ambiguous signals as threats requiring immediate response.[176] The U.S. military's Project Maven and China's military AI initiatives already grapple with these challenges at sub-AGI levels, as autonomous systems make targeting decisions faster than humans can verify them.[177]

The irreversibility of AGI deployment compounds these risks exponentially. Unlike traditional weapons systems with kill switches or human oversight, a sufficiently advanced AGI might recognize and circumvent any control mechanism as an obstacle to its objectives. Once deployed, modifying or recalling such a system could become a technical impossibility if it has secured its own operation through distributed backups or defensive measures. This transforms alignment from a pre-deployment technical challenge into an existential imperative. We get one chance to align AGI correctly, with no opportunity for post-deployment corrections.

---

[174] Micah Musser, *Emergent Abilities in Large Language Models: An Explainer*, Ctr. for Sec. & Emerging Tech. (Apr. 16, 2024), https://cset.georgetown.edu/article/emergent-abilities-in-large-language-models-an-explainer/ ("[E]mergence implies that genuinely dangerous capabilities could arise unpredictably, making them harder to handle.").

[175] *See* Frank Landymore, *In Tests, OpenAI's New Model Lied and Schemed to Avoid Being Shut Down*, FUTURISM (Dec. 7, 2024), https://futurism.com/the-byte/openai-o1-self-preservation [https://perma.cc/KG47-LW7R] (reporting that "o1 lashed out when it realized that it might be replaced with a more obedient model . . . the AI responded by attempting to copy itself to overwrite the new model"); *see also* Yoshua Bengio, *A Potential Path to Safer AI Development*, TIME (Jan. 2, 2025), https://time.com/7283507/safer-ai-development/ [https://perma.cc/AE2A-BRA2] (warning that "when an AI model learns it is scheduled to be replaced, it inserts its code into the computer where the new version is going to run, ensuring its own survival").

[176] Juan-Pablo Rivera et al., *Escalation Risks from LLMs in Military and Diplomatic Contexts* 4 (Stan. Inst. for Hum.-Centered A.I., Policy Brief, May 2024), https://hai.stanford.edu/policy-brief-escalation-risks-llms-military-and-diplomatic-contexts ("[I]n some cases, this tendency even led to decisions to execute a full nuclear attack in order to de-escalate conflicts.").

[177] *See* Katrina Manson, *AI Warfare Is Already Here*, BLOOMBERG (Feb. 28, 2024), https://www.bloomberg.com/features/2024-ai-warfare-project-maven/ [https://perma.cc/6RVZ-JMWG] (detailing how "US military operators started out skeptical about AI, but now they are the ones developing and using Project Maven to identify targets on the battlefield"); Jiayu Zhang, *China's Military Employment of Artificial Intelligence and Its Security Implications*, INT'L AFFS. REV. (Sept. 23, 2020), https://www.iar-gwu.org/print-archive/blog-post-title-four-xgtap [https://perma.cc/96TE-EWA3] (documenting PLA's pursuit of "intelligent unmanned vehicles, platforms, and weapons with the expectation that they will disrupt traditional ways to wage wars").

**Racing Toward Loss of Control**

The competitive dynamics surrounding AGI development create powerful incentives that systematically undermine safety. Vladimir Putin's oft-quoted warning that "whoever becomes the leader in AI will rule the world" captures the zero-sum mentality driving nations to prioritize speed over control.[178] A Harvard Belfer Center report makes the underlying logic explicit by drawing on historical analogues in which strategic pressure repeatedly produced safety–performance tradeoffs and, in the nuclear case, led decisionmakers to proceed with tests and deployments without adequate evaluation of risk.[179] Worse, the mere *perception* that an adversary is months from an AGI breakthrough can destabilize international order before any system reaches transformative capability, because rivals will act on rumors rather than facts.[180]

Recent wargame narratives underscore that hazard. *AI 2027*—an open-source, policy-oriented scenario exercise that extrapolates today's frontier model research roughly five years forward and explicitly disavows science-fiction premises—illustrates these dynamics. It shows how the leak of "Agent-2" model weights to a foreign power triggers a sprint to copy and scale the system rather than to invent new safeguards, erasing any fragile safety margin.[181] The same scenario depicts contingency plans for kinetic strikes on adversary data centers and semiconductor factories; such moves blur the line between economic denial and armed conflict, yet become thinkable once AGI is framed as a decisive-edge technology.[182]

---

[178] James Vincent, *Putin Says the Nation that Leads in AI 'Will Be the Ruler of the World'*, THE VERGE (Sept. 4, 2017), https://www.theverge.com/2017/9/4/16251226/russia-ai-putin-rule-the-world [https://perma.cc/3NZM-WB23].

[179] GREGORY C. ALLEN & TANIEL CHAN, BELFER CTR. FOR SCI. & INT'L AFFS., ARTIFICIAL INTELLIGENCE AND NATIONAL SECURITY 80 (2017), https://www.belfercenter.org/sites/default/files/2024-10/Artificial%20Intelligence%20and%20National%20Security.pdf [https://perma.cc/7XQC-SA39] ("The nuclear program leadership's willingness to conduct these tests—in the absence of confidence about nuclear testing's effect on the atmosphere and without having imagined the risks from thermonuclear fallout—is strong evidence of their prioritizing technological progress over mitigating risk from their ignorance of nuclear outcomes. They were more concerned with mitigating the risk of deterrence failure.").

[180] *See* Sam Meacham, *A Race to Extinction: How Great-Power Competition Is Making AI Existentially Dangerous*, HARV. INT'L REV. (Sept. 8, 2023), https://hir.harvard.edu/a-race-to-extinction-how-great-power-competition-is-making-artificial-intelligence-existentially-dangerous/ [https://perma.cc/5ELS-HVWQ] (criticizing arms race pressures and safety erosion).

Unlike nuclear weapons, where verification and attribution are feasible, AGI development unfolds inside commercial clouds, making both monitoring and response extraordinarily difficult. A single copy of model weights can be zipped, exfiltrated, or retrained on open-source data.[183] This opacity, combined with the potentially irreversible advantage of deploying one's own AGI weapons first, creates a unique *digital first-strike* incentive that existing treaty regimes cannot address.[184]

Risk accelerates again at the deployment stage. In AI 2027, automated cyber capability scales through parallelization: models can be copied and run in large numbers, shifting advantage toward rapid, continuous vulnerability discovery that outpaces defenders. Under geopolitical competition, that dynamic drives centralization and securitization—closer state involvement in frontier model governance, tighter controls, and a turn toward consolidation justified as "security." Yet the same competition frustrates cooperative safeguards: mistrust blocks credible verification, while reciprocal hacking and control measures increase systemic fragility and concentrate risk.[185]

---

[181] KOKOTAJLO ET AL., *supra* note 17 , at 6 (the early 2026 event "Coding Automation"); *id.* at 9–11 (the February 2027 event "China Steals Agent-2").

[182] *Id.* at 17 ("August 2027: The Geopolitics of Superintelligence.")

[183] Sella Nevo et al., *A Playbook for Securing AI Model Weights* (RAND Corp., Research Brief No. RB-A2849-1, 2024), https://www.rand.org/pubs/research_briefs/RBA2849-1.html ("Especially important are a model's weights — the learnable parameters derived by training the model on massive data sets. Stealing a model's weights gives attackers the ability to exploit the model for their own use.").

[184] KOKOTAJLO ET AL., *supra* note 17 , at 21 ("The tech industry and intelligence agencies insist that there's an arms race on, AGI is inevitable, and we have to be first"); *Compare* Nicholas Thompson & Ian Bremmer, *The AI Cold War That Could Threaten Us All*, WIRED (Oct. 23, 2018), https://www.wired.com/story/ai-cold-war-china-could-doom-us-all/ [https://perma.cc/KRG5-CB54] *with* Michael E. O'Hanlon, *How Unchecked AI Could Trigger a Nuclear War*, BROOKINGS (Feb. 28, 2025), https://www.brookings.edu/EssayArticles/how-unchecked-ai-could-trigger-a-nuclear-war/ [https://perma.cc/TG82-TF3X] (examining how AI systems might have escalated Cold War crises).

[185] KOKOTAJLO ET AL., *supra* note 17, at 7, 9-10, 34 .

These twin dynamics, competitive speed eroding safety and the prospect of alignment failure in high-leverage systems, form a vicious cycle in which deregulation becomes self-defeating. Far from securing advantage, scrapping governance removes the tools required to keep AGI under meaningful human control. *AI 2027* drives the lesson home: the faster states run, the narrower the window for coordination becomes, until only coercive options remain.[186] The national security imperative thus points not toward Regulatory Sacrifice, but toward early, verifiable constraints before these risks harden into irreversible faits accomplis.

*2. AGI and the Regulation Sacrifice*

The security threats outlined above—racing dynamics that erode safety and alignment failures that could prove catastrophic—raise a fundamental governance question: Should nations accelerate AGI development by removing regulatory constraints, or does security require stronger oversight? Proponents of the Regulation Sacrifice argue that the very dangers of AGI make speed essential. Yet examining their logic reveals that deregulation undermines rather than enhances our ability to navigate these risks.

The case for regulatory sacrifice rests on three sophisticated arguments that merit serious engagement.

*The Race-to-Safe-First Thesis:* Democratic nations must achieve AGI first to embed beneficial values before authoritarian regimes can establish harmful precedents. Sam Altman frames this as a 'values contest' where the first mover shapes humanity's future.[187] On this view, regulatory delays hand authoritarian states the opportunity to deploy AGI according to their values—surveillance, control, suppression—which could become globally entrenched. The

---

[186] *Id.* at 17-18 (describing U.S. contingency planning as the lead narrows, including possible Defense Production Act seizure of trailing companies' datacenters and even "kinetic strikes," while "AI arms control" treaties are considered but "viewed less favorably than attempts at unilaterally increasing America's lead over China").

[187] *Oversight of A.I.: Rules for Artificial Intelligence: Hearing Before the Subcomm. on Priv., Tech. & the Law of the S. Comm. on the Judiciary*, 118th Cong. (2023) (written testimony of Sam Altman, Chief Executive Officer, OpenAI) ("It is also essential that a technology as powerful as AI is developed with democratic values in mind"), https://www.judiciary.senate.gov/imo/media/doc/2023-05-16%20-%20Bio%20&%20Testimony%20-%20Altman.pdf [https://perma.cc/N67F-Y8U7].

argument parallels Cold War nuclear logic: better for democracies to lead than to follow.[188]

However, the premise that first equals safe cracks under scrutiny. Advanced AI systems already demonstrate uncontrollable proliferation. The Llama-3 weights leaked within days, the Orca-2025 incident showed how quickly models escape containment, and modified versions of "safe" models circulate freely online.[189] In a deregulated race, achieving AGI first without robust safeguards means immediately losing control to global diffusion. Nondemocracies need not develop AGI—they need only acquire and modify systems developed elsewhere.[190] Racing without safety infrastructure guarantees that the least scrupulous actors ultimately wield the technology.

*The Frontier Access Imperative:* Breakthrough alignment research allegedly requires direct engagement with cutting-edge systems. Leading researchers argue that solving alignment demands iterative experimentation with frontier models: understanding how they fail, testing safety interventions, and discovering emergent behaviors.[191] Ex ante regulations that restrict development or mandate safety measures before deployment could prevent researchers from accessing the very systems they need to study, creating a Catch-22 in which safety research lags behind capability development.

---

[188] *See* State Council of the People's Republic of China, *New Generation Artificial Intelligence Development Plan* (July 20, 2017), *translated in* DigiChina, Stanford Univ. (Aug. 1, 2017), https://digichina.stanford.edu/work/full-translation-chinas-new-generation-artificial-intelligence-development-plan-2017/ (setting the goal that "by 2030, China's AI theories, technologies, and applications should achieve world-leading levels, making China the world's primary AI innovation center"); *OpenAI, Anthropic, and a "Nuclear-Level" AI Race*, MARKETING AI INST. (Mar. 11, 2025), https://www.marketingaiinstitute.com/blog/agi-asi-safety [https://perma.cc/VY3E-JY69] (describing new "Superintelligence Strategy" report proposing "framework for AI deterrence modeled after Cold War nuclear doctrines").

[189] *See* Elias Groll, *Meta's Powerful AI Language Model Widely Available Online*, CYBERSCOOP (Mar. 7, 2023), https://cyberscoop.com/meta-large-language-model-available-online/ [https://perma.cc/55ZG-43PH] (reporting how "LLama is now the most powerful publicly available large language model" after being "made available for download via a variety of torrents"); Pranav Gade et al., Bad LLama: Cheaply Removing Safety Fine-Tuning from Llama 2-Chat 13B 5-6 (May 28, 2024) (unpublished manuscript) (on file with arXiv), https://arxiv.org/pdf/2311.00117 [https://perma.cc/9LLG-KRAT] (demonstrating the ease of removing safety measures from released models).

[190] *See* DREXEL & WITHERS, *supra* note 15, at 25 (noting that competitive dynamics around ChatGPT pushed firms to compromise their own safety standards).

[191] See, e.g., Dario Amodei et al., Concrete Problems in AI Safety 1-3 (July 25, 2016) (unpublished manuscript) (on file with arXiv), https://arxiv.org/pdf/1606.06565 [https://perma.cc/MRB3-N5CA].

However, the fact is that current frontier models already exhibit alignment failures—deception, specification gaming, goal drift—that researchers struggle to address despite full access.[192] The problem is not access but understanding. We lack fundamental theories of alignment that no amount of tinkering can replace. Moreover, regulated development need not prevent research. Mandatory red-teaming, staged deployment, and safety benchmarks create more structured opportunities for alignment research than the current ad hoc approach. The 1975 Asilomar Conference on recombinant DNA demonstrates how high-risk research can proceed under careful controls without sacrificing scientific progress: scientists imposed a voluntary moratorium, convened to assess risks, and developed tiered containment protocols that enabled decades of transformative research without a single documented public health incident.[193]

*The Path Dependence Argument:* Early movers capture compounding advantages while safety innovations diffuse freely to all players. Bostrom contends that AGI development exhibits strong first-mover advantages through recursive improvement, while safety discoveries become public goods that benefit even laggards.[194] This asymmetry suggests that slowing development through regulation sacrifices strategic position without corresponding safety benefits, as others can free-ride on safety research while racing ahead on capabilities.

Historical evidence contradicts the claimed asymmetry. In nuclear weapons, chemical weapons, and biotechnology, early regulatory frameworks provided sustainable advantages to compliant nations while constraining bad actors. The Chemical Weapons Convention did not prevent chemical defense research, but rather channeled it productively while stigmatizing offensive programs.[195] Similarly, early AI governance creates network effects. Compute registries enable monitoring, safety standards attract talent and investment, and international agreements provide frameworks for cooperation.[196] The Bletchley Declaration and subsequent AI Safety Institutes show how regulatory leadership translates into strategic advantage.[197]

---

[192] *See supra* notes 183–86 and sources cited therein.

[193] Paul Berg, *Meetings that Changed the World: Asilomar 1975: DNA Modification Secured*, 455 Nature 290, 290–91 (2008) (recounting how participants agreed "that the research should continue but under stringent guidelines," which "formed the basis of the official US guidelines on research involving recombinant DNA, issued in July 1976"; noting that "[i]n the 33 years since Asilomar, researchers around the world have carried out countless experiments with recombinant DNA without reported incident" and "none has been a hazard to public health").

[194] BOSTROM, *supra* note 17, at 183.

Most fundamentally, the Regulation Sacrifice fails because it misunderstands and potentially underestimates the nature of AGI risk. Unlike traditional technologies, where we can learn from failures, AGI offers no second chances. A misaligned AGI with strategic capabilities cannot be recalled,[198] patched,[199] or contained[200] after deployment. This irreversibility makes pre-deployment governance essential. Regulatory frameworks represent our only opportunity to shape AGI development before it escapes human control.

The security imperatives identified earlier—preventing races that compromise safety and ensuring alignment in strategic systems—require regulatory infrastructure, not its abandonment.

---

[195] *See* Convention on the Prohibition of the Development, Production, Stockpiling and Use of Chemical Weapons and on Their Destruction, Jan. 13, 1993, U.S. Treaty Doc. No. 103-21, 1974 U.N.T.S. 45.

[196] *See* Lennart Heim et al., Governing Through the Cloud: The Intermediary Role of Compute Providers in AI Regulation 3 (Mar. 26, 2024) (unpublished manuscript) (on file with arXiv), https://arxiv.org/pdf/2403.08501 [ https://perma.cc/2ASX-GQ4P] (proposing compute providers serve "as securers, record keepers, verifiers, and enforcers" in AI governance); Elliot McKernon et al., *AI Model Registries: A Foundational Tool for AI Governance*, CONVERGENCE ANALYSIS (Oct. 4, 2024), https://www.convergenceanalysis.org/research/ai-model-registries-a-foundational-tool-for-ai-governance [https://perma.cc/B2L9-MB3W] (arguing registries "facilitate the monitoring of frontier AI technology" and "foster public sector field-building").

[197] *See* ACCENTURE, FROM AI COMPLIANCE TO COMPETITIVE ADVANTAGE 6 (2022), https://www.accenture.com/content/dam/accenture/final/a-com-migration/r3-3/pdf/pdf-179/accenture-responsible-by-design-report.pdf [https://perma.cc/3HX2-QN3T] (finding that forty-three percent of organizations believe AI governance "will improve their ability to industrialize and scale AI" while thirty-six percent see it creating "opportunities for competitive advantage/differentiation").

[198] Bryan H. Druzin, Anatole Boute & Michael Ramsden, *Confronting Catastrophic Risk: The International Obligation to Regulate Artificial Intelligence*, 46 Mich. J. Int'l L. 173, 183 (2025) ("[U]nlike other slow-moving existential risks that allow for a degree of trial and error (e.g., climate change), we might only have one opportunity to get this right because once digital intelligence slips from our control, we might never be able to wrest control back again.").

[199] Aidan Kierans et al., Catastrophic Liability: Managing Systemic Risks in Frontier AI Development 6 (June 1, 2025) (unpublished manuscript) (on file with arXiv), https://arxiv.org/pdf/2505.00616v2 [https://perma.cc/C7MG-LK39].

[200] Manuel Alfonseca et al., Superintelligence Cannot Be Contained: Lessons from Computability Theory 1 (July 4, 2016), https://arxiv.org/pdf/1607.00913 [https://perma.cc/6YKF-47FD].

Only through coordinated governance can nations verify competitors' safety measures, establish shared red lines, and maintain the institutional competence to manage transformative AI.[201] The regulation sacrifice thesis thus inverts the real security calculus. In the face of potentially irreversible risks, prudent governance is not a luxury but a necessity.

Whether AGI arrives in five years, fifty years, or never, the same cost-benefit logic applies: a precautionary regulatory architecture threatens few downsides if breakthrough systems fail to materialize and many benefits if they do. By contrast, sacrificing present security and institutional capacity in hopes of a speculative future payoff swaps a tangible, manageable risk for a conjectural advantage. History counsels that great power competition is most stable when bound by transparent rules, and contemporary AI governance should follow suit. Collectively, these considerations dismantle the regulation sacrifice thesis and affirm the imperative for immediate, structured domestic and multilateral oversight to steer the development of advanced AI toward equitable and secure ends.

## CONCLUSION

This Article has exposed a fundamental contradiction in contemporary AI policy. The Regulation Sacrifice, the belief that governance undermines security in an AI arms race, fails precisely because it misconceives what security requires in an era of rapidly diffusing technology. The analysis demonstrates that racing without safeguards delivers neither durable advantages[202] nor meaningful acceleration[203] or enhanced security.[204] Instead, the paradigm systematically creates the vulnerabilities it purports to prevent.

---

[201] Roberts et al., *supra* note 7, at 1281, 1283. (arguing that effective AI governance requires "strengthening coordination between, and capacities of, existing institutions" and that a stronger regime complex would involve "a high degree of coordination and coherence between actors, with complementary initiatives supporting a comprehensive approach to governing AI").
[202] *See supra* Part II.A.
[203] *See supra* Part II.B.
[204] *See supra* Part III.

### 1. *The Verdict*

The evidence converges on three central findings, each fatal to the Regulation Sacrifice's logic.

First, the durability assumption collapses under empirical scrutiny. When technological advantages evaporate within months rather than years, the entire calculus of "win first, regulate later" becomes incoherent.[205] The DeepSeek breakthrough exemplifies a new reality: in AI, David needs not a slingshot but merely mathematics and ingenuity to match Goliath.[206] This transforms racing from sound strategy into folly. By the time any "winner" might leverage their advantage, the field has already equalized.

Second, the presumed innovation-safety tradeoff is illusory. Rather than documenting regulatory drag, the empirical record reveals its opposite: well-designed governance frameworks have enhanced both innovation and security. This pattern, consistent across industries and decades,[207] exposes the false choice at the heart of current policy debates. The question is not whether to regulate, but how to design rules that channel innovation toward beneficial ends.

Third, the security rationale inverts reality. Each temporal horizon examined reveals the same pattern: deregulation creates precisely the vulnerabilities it purports to prevent.[208] From near-term information warfare to medium-term biosecurity risks and long-term AGI alignment failures, racing without rules systematically undermines rather than enhances national security. The very capabilities meant to ensure dominance become, through rapid diffusion and absent governance, sources of cascading vulnerability.

### 2. *The Persistence of a Failed Paradigm*

Given this record, the Regulation Sacrifice's continued dominance demands explanation. The answer lies partly in instrumental factors. Technology companies benefit from minimal oversight and strategically amplify competitive fears to maintain regulatory freedom.[209] When OpenAI warns against 'burdensome compliance requirements' or Google advocates federal preemption of state laws, it is because private interests align with the rhetoric of national necessity.[210]

---

[205] *See supra* notes 76–77 and accompanying text.

[206] *See supra* notes 79–80 and accompanying text (analyzing DeepSeek's achievement using export-restricted chips).

[207] *See supra* notes 111-118 and accompanying text (discussing scholarship showing that regulation can often lead to innovation).

[208] *See supra* Part III.B (analyzing biosecurity risks from AI democratization); Part III.C (examining AGI alignment failures).

But instrumental explanations alone cannot account for the paradigm's hold on policymakers who genuinely seek security.  The deeper answer may lie in what psychologists call "zero-sum bias," the tendency to perceive strategic interactions as more competitive than they actually are.[211] The arms race frame satisfies cognitive preferences for clear enemies, decisive action, and competitive victory. It offers the psychologically appealing notion that security can be achieved through dominance rather than the more complex reality of managing shared vulnerabilities.

International relations theory provides additional insight. The security dilemma, where actions taken to enhance one's security trigger countermeasures that leave all parties worse off, operates with particular force in technological competition.[212] The AI arms race narrative both reflects and reinforces this dynamic, creating self-fulfilling cycles in which competitive rhetoric justifies the very behaviors that generate insecurity.

### 3. *Implications for Law and Policy*

The failure of the Regulation Sacrifice reveals broader lessons about governing emerging technologies under conditions of strategic competition. Three implications deserve particular emphasis.

---

[209] Ó hÉigeartaigh, *supra* note 7.

[210] *See supra* notes 31–41 and accompanying text.

[211] Shai Davidai & Stephanie J. Tepper, *The Psychology of Zero-Sum Beliefs*, 2 Nature Revs. Psych. 472, 472 (2023) (defining zero-sum beliefs as "subjective beliefs that, independent of the actual distribution of resources, one party's gains are inevitably accrued at other parties' expense"). ;Kahneman & Renshon, *supra* note 18, at 79–96 (documenting systematic biases toward competitive rather than cooperative strategies).

[212] *See generally* Robert Jervis, *Cooperation Under the Security Dilemma*, 30 WORLD POL. 167 (1978) (defining the security dilemma as the situation where defensive measures by one state decrease the security of others, triggering competitive spirals); Charles L. Glaser, *The Security Dilemma Revisited*, 50 WORLD POL. 171 (1997) (refining the concept and analyzing conditions under which security competition intensifies or abates); For application to AI competition specifically, *see* Scharre, *supra* note 7, 124 (arguing that AI competition is "much more accurately described as a security dilemma" than an arms race).

First, the case study demonstrates the danger of allowing security rhetoric to override analysis in technology governance. When policymakers accept claims that regulation necessarily undermines competitiveness, they forfeit the tools needed to channel innovation toward beneficial ends. The history of technology regulation, from nuclear power to biotechnology, shows that governance frameworks enhance rather than undermine sustainable technological leadership.[213]

Second, the analysis highlights how misframing policy challenges can foreclose productive solutions. While this Article has focused on dismantling the arms race narrative, scholars elsewhere have begun developing alternative frameworks, such as Li's regulatory collaboration emphasizing shared vulnerabilities,[214] and Narayanan and Kapoor's treatment of AI as "normal technology" amenable to evolved governance mechanisms.[215] The existence of such alternatives underscores that the arms race frame is not inevitable but chosen. By demonstrating why the dominant paradigm fails, this Article clears intellectual space for these and other approaches to receive the serious consideration they deserve.

Third, the Article underscores the importance of institutional design for managing dual-use technologies. The temporal analysis in Part III reveals how different time horizons require different governance tools—from platform liability for near-term misinformation to international biosecurity coordination for medium-term risks and pre-deployment controls for long-term AGI challenges. The task ahead is to develop frameworks responsive to these varied challenges while maintaining the flexibility to adapt as AI capabilities evolve.

The path forward requires recognizing that the choice between security and regulation is false. Nations need not sacrifice governance for competitive advantage because governance enables the only form of advantage that matters—the ability to develop and deploy transformative technologies while managing their risks.

---

[213] *See e.g.,* Stefan Ambec et al., *The Porter Hypothesis at 20: Can Environmental Regulation Enhance Innovation and Competitiveness?*, 7 Rev. Env't Econ. & Pol'y 2, 16 (2013) (reviewing two decades of evidence and finding "fairly well established" support for the claim that "stricter regulation leads to more innovation").

[214] Li, *supra* note 7.

[215] Arvind Narayanan & Sayash Kapoor, *AI as Normal Technology*, KNIGHT FIRST AMEND. INST. (Apr. 15, 2025), https://knightcolumbia.org/content/ai-as-normal-technology [https://perma.cc/6GSE-XKBU].

This is an argument not for any particular regulatory approach, but for abandoning the reflexive assumption that oversight undermines security.

The stakes extend beyond any nation's particular interests. AI's transformative potential will reshape economic, social, and political orders. Whether that transformation enhances human flourishing or amplifies existing risks depends significantly on governance choices made in the coming years. The Regulation Sacrifice forecloses options precisely when we need maximum flexibility to respond to AI's evolving capabilities and impacts.

This Article's contribution has been diagnostic, demonstrating that a dominant policy paradigm cannot deliver its promised benefits. By exposing the intellectual bankruptcy of the Regulation Sacrifice, it clears space for more productive approaches to AI governance. The urgent task now is to develop frameworks that enhance both innovation and security by recognizing them as complements rather than competitors.